\documentclass[aps,prd,nofootinbib,onecolumn,amsmath,amssymb,superscriptaddress,floatfix,scrartcl]{revtex4-1}

\usepackage{amssymb}  
\usepackage{color,soul} 
\usepackage[utf8]{inputenc}
\usepackage{graphicx}  
\usepackage[tight]{subfigure}  
\usepackage{mathrsfs}
\usepackage{bbm}
\usepackage[pdftex,pdfusetitle,bookmarks=true,colorlinks=true,citecolor=blue,urlcolor=blue,linkcolor=magenta]{hyperref}
\usepackage{hypcap}
\usepackage{changes}
\usepackage{xcolor}

\usepackage{graphicx}

\newcommand{\inc}{\textrm{inc}}

\begin{document}
\title{Effective holographic theory of charge density waves}

\author{Andrea Amoretti}   
\email{andrea.amoretti@ulb.ac.be} 
\affiliation{Physique Th\'{e}orique et Math\'{e}matique Universit\'{e} Libre de Bruxelles, C.P. 231, 1050 Brussels, Belgium}

\author{Daniel Are\'an}
\email{daniel.arean@fc.up.pt}
\affiliation{Centro de F\'\i sica do Porto, Departamento de F\'\i sica da
Universidade do Porto, Rua do Campo Alegre 687, 4169-007 Porto, Portugal}

\author{Blaise Gout\'eraux}
\email{blaise.gouteraux@su.se}
\affiliation{Nordita, KTH Royal Institute of Technology and Stockholm University, Roslagstullsbacken 23, SE-106 91 Stockholm, Sweden}

\author{Daniele Musso}
\email{daniele.musso@usc.es}
\affiliation{Departamento  de  F\'{i}sica  de  Part\'{i}culas,  Universidade  de  Santiago  de  Compostela  and  Instituto  Galego  de  F\'{i}sica  de  Altas  Enerx\'{i}as  (IGFAE),  E-15782,  Santiago  de  Compostela, Spain.}

\begin{abstract}
We use Gauge/Gravity duality to write down an effective low energy holographic theory of charge density waves. 
We consider a simple gravity model which breaks translations spontaneously  in the dual field theory in a homogeneous manner, capturing the low energy dynamics of phonons coupled to conserved currents. We first focus on the leading two-derivative action, which leads to excited states with non-zero strain. We show that including subleading quartic derivative terms leads to dynamical instabilities of AdS$_2$ translation invariant states and to stable phases breaking translations spontaneously.
We compute analytically the real part of the electric conductivity. The model allows to construct Lifshitz-like hyperscaling violating quantum critical ground states breaking translations spontaneously. At these critical points, the real part of the dc conductivity can be metallic or insulating.
\end{abstract}

\maketitle   
\tableofcontents

\section{Introduction}
Many condensed matter systems are described by non-relativistic effective Hamiltonians, due to the breaking of translations by the underlying ionic lattice. The standard approach to consider the effects of the lattice on the electronic subsystem is to treat it as an external explicit source of momentum relaxation.  In many cases of interest (e.g. high $T_c$ superconductors), the strongly coupled electronic fluid also tends to break translations spontaneously, developing spatial modulations with a periodicity incommensurate to that of the ionic lattice. This includes the formation of charge density wave (CDW) and spin density wave orders. Constructing effective strongly interacting field theories for spontaneous symmetry breaking of translations is therefore extremely relevant to understand the behaviour of these systems.

Reliable theoretical descriptions of these phenomena are challenging at strong coupling. Field theory approaches exist (see \cite{2017arXiv170308172L} for a review and references therein), which typically couple a gapless, critical boson to a Fermi surface. In $d=2$, such theories are strongly coupled in the IR and can only be analyzed in certain limits. Alternatively, long wavelength Effective Field Theories (EFT) of CDWs \cite{chaikin_lubensky_1995, RevModPhys.60.1129,Delacretaz:2017zxd} have been written down. However, they are limited to the low frequency, low wavevector regime and do not provide a microscopic description of the ground state. 

Gauge/Gravity duality offers an intermediate approach \cite{Ammon:2015wua,Zaanen:2015oix,Hartnoll:2016apf} by mapping the problem to a weakly-coupled, classical theory of gravity. Top-down constructions correspond to specific field theory duals, but are usually less tractable, and only exist for particular values of the low energy couplings, e.g. \cite{Donos:2011qt,Donos:2012yu,Jokela:2014dba,Jokela:2016xuy,Faedo:2017aoe}.
Bottom-up models \cite{Charmousis:2010zz} offer less control over the microscopic content of the dual field theory, but allow to scan more easily for interesting phenomenology.
Dynamical instabilities of translation-invariant holographic states towards phases with spatial modulation have been thoroughly characterized \cite{Ooguri:2010kt,Donos:2011bh,Donos:2011qt,Donos:2011ff,Cremonini:2012ir,Cremonini:2013epa,Donos:2013gda,Gouteraux:2016arz}. The corresponding spatially modulated solutions in a variety of holographic setups have also been constructed \cite{Donos:2012wi,Donos:2013wia,Withers:2013loa,Jokela:2014dba,Withers:2014sja,Cremonini:2016rbd,Jokela:2016xuy,Cremonini:2017usb,Cai:2017qdz}, dual to various kinds of density waves. Progress on the understanding of their transport properties has been slower. This is in part because most of the work has focussed either on models based on the homogeneous Bianchi VII$_0$ subgroup \cite{Andrade:2017cnc}, which is special to five-dimensional bulks and leads to fairly complicated solutions; on probe brane models, where it can be hard to understand the precise consequences of freezing the metric degrees of freedom or taking the probe limit \cite{Jokela:2016xuy,Jokela:2017ltu}; or on inhomogeneous geometries \cite{Andrade:2017ghg}, which are more realistic but for the most part can only be studied numerically (though see \cite{Donos:2017gej,Donos:2017ihe}). In contrast, much has been understood about explicit translation symmetry breaking by thoroughly studying conceptually simpler homogeneous models, based on massive gravity, Q-lattices or Stückelberg scalars \cite{Vegh:2013sk,Davison:2013jba,Blake:2013bqa,Andrade:2013gsa,Donos:2013eha,Donos:2014uba,Gouteraux:2014hca,Donos:2014cya,Amoretti:2014zha,Amoretti:2014mma,Kim:2014bza,Davison:2014lua,Davison:2015bea}.

In the present paper, we explore a class of holographic models akin to those of \cite{Andrade:2013gsa,Donos:2013eha,Donos:2014uba,Gouteraux:2014hca}, but where translations are broken spontaneously \cite{Amoretti:2016bxs} rather than explicitly. As in the explicit case, the simplicity of the model allows us to go quite far in understanding its properties analytically. Instead of considering inhomogeneous states that break spontaneously the translations and later taking the long wavelength limit, we directly describe the coupled dynamics of conserved densities (energy, momentum, density) and Nambu-Goldstone modes. Such Goldstones are the fundamental constituents of the low energy EFT description and are essential to the dynamics \cite{PhysRev.177.2239,PhysRev.177.2247}. We show that our model reproduces correctly various aspects of the EFT of CDW states \cite{chaikin_lubensky_1995, RevModPhys.60.1129}, including transport properties.

The plan of this paper is as follows. In section \ref{section:HomHolSTSB} we construct the effective holographic theory, explain how translations are broken spontaneously in the boundary dual field theory and compute the holographic on-shell action at quadratic order in fluctuations and one-point functions. We first focus on a two-derivative model, which contains unstable phases with non-zero strain. We explain how stable phases can be captured when quartic-derivative terms are included in the EFT, and find they source dynamical instabilities of translation invariant phases. We show explicitly that the equilibrium holographic stress-energy tensor agrees with that of an isotropic crystal.
In section \ref{subsection:cond} we compute analytically the low frequency limit of the real part of the electric conductivity using Kubo formul\ae. 
In section \ref{section:QCPs} we construct quantum critical CDW phases, which can have non-trivial Lifshitz dynamical and hyperscaling violating exponents \cite{Gouteraux:2014hca}. By combining with the results obtained in section \ref{subsection:cond}, we give a prediction for the low temperature scaling of the real part of the dc conductivity.
Next, in section \ref{section:BHs} we construct numerically homogeneous black holes dual to finite temperature states breaking translations spontaneously with non-zero strain, with either finite or vanishing entropy at zero temperature. 
We conclude with some further discussion and future directions in section \ref{section:outlook}.

In a companion paper \cite{Amoretti:2017axe}, we explain the relevance of our results to charge transport at a weakly pinned CDW quantum critical point and connections to transport in cuprate high $T_c$ superconductors.

\noindent\emph{Note added:} As this work was in the final stages, we became aware of \cite{Alberte:2017oqx} which also studies a homogeneous model of spontaneous translation symmetry breaking in holographic massive gravity, following earlier work in \cite{Alberte:2017cch}.

\noindent\emph{Second note added:} After this work appeared as a preprint, \cite{Donos:2018kkm} emphasized how considering thermodynamically stable phases affects the incoherent conductivity. The new version of this work reflects this improved understanding.


\section{Homogeneous spontaneous translation symmetry breaking \label{section:HomHolSTSB}}

We first present our effective holographic theory of long wavelength dynamics of CDW states. Then we explain how including subleading, quartic derivative terms triggers dynamical instabilities of translation-invariant states towards phases breaking translations spontaneously.

\subsection{Two-derivative model: excited phases with non-zero strain}
\subsubsection{Setup \label{subsection:setup}}

In a CDW state, the charge density is expressed as $\rho(x,t)=\rho_0+\rho_1(x,t)\cos(k x+\Psi(x,t))$ \cite{ RevModPhys.60.1129}.
In the EFT, the order parameter is described by means of a complex scalar \cite{2015RvMP...87..457F} whose phase is expanded at linear order around equilibrium as $k x+\Psi(x,t)$. $\rho_1$ models amplitude fluctuations of the order parameter, $\Psi$ phase fluctuations. The latter are gapless modes, ie the phonons of spontaneous translation symmetry breaking. Above $T_c$, both type of fluctuations are expected and part of the EFT. However, below $T_c$, the long wavelength dynamics is described by the interplay between conserved quantities and the phonons \cite{chaikin_lubensky_1995}.

This motivates us to consider a generalized complex scalar action \cite{Donos:2013eha}
\begin{equation}\label{actionqlattice}
S=\int d^{d+2}x\,\sqrt{-g}\left[R-Y_\Phi(|\Phi_I|)\delta^{IJ}\partial\Phi_I\partial\Phi_J^*-\frac14Z_\Phi(|\Phi_I|)F^2-V_\Phi(|\Phi_I|)\right],
\end{equation}
where $I,J=1\dots d$ run over the spatial coordinates of the boundary. This gravitational model is dual to a CFT deformed by complex scalar operators. 

The model has a global U(1) symmetry $\Phi_I\to\Phi_I \exp(i c_I)$ where $c_I$ is just a constant, which can also be viewed as a shift symmetry of the phase of the complex field. Following \cite{Donos:2013eha}, we adopt the following background Ansatz for the complex scalars $\Phi_I = \varphi(r) \exp(i k \delta_{Ii} x^i)$. It breaks both spatial translations $x^i\to x^i+a^i$ and the shift symmetry, but preserves a diagonal subgroup \cite{Nicolis:2013lma}. Thus, it is consistent to assume the other fields in the bulk not to depend on $x^i$, which considerably simplifies solving the model.\footnote{Isotropy also follows from a similar breaking of internal rotations of the $\Phi_I$ and spacetime rotations down to a diagonal subgroup.} 
From now on, we no longer need to distinguish between $i$ and $I$ indices.
The (real) scalar $\varphi$ has the following asymptotic expansion at the Anti de Sitter boundary $r\to0$ 
\begin{equation}\label{eq:phiBCmod}
\varphi(r\to0)=\varphi_{(s)}r^{d+1-\Delta}+\varphi_{(v)}r^{\Delta}+\dots\,,\qquad m_\Phi^2=\Delta(\Delta-d-1)\,,
\end{equation}
where $m_\Phi$ is the mass of the scalars $\Phi_I$ (which we take to be the same for all $\Phi_I$, for simplicity), which is related to $V_\Phi''(0)$ in the usual fashion. If $\varphi_{(s)}\neq0$, translations are broken explicitly by the background, while if $\varphi_{(s)}=0$, the breaking is spontaneous. Our interest is in the second case.

In this work, we are mostly interested in linear response at zero wavevector, so it is enough to consider linear fluctuations around the background. The scalar fluctuations which enter the calculation of the conductivities and preserve the homogeneity of the eoms are
\begin{equation}
\delta\Phi_I = \delta\varphi(r)e^{i k \delta_{Ii} x^i}e^{-i\omega t}
\end{equation}
and can be rewritten in a `polar' decomposition
\begin{equation}
\Phi_I +\delta\Phi_I= \varphi(r)e^{i k \delta_{Ii} x^i+i\delta\psi_I(r,t)}\,,\quad \delta\psi_I(r,t)=-i\frac{\delta\varphi(r)}{\varphi(r)}e^{-i\omega t}\,,
\end{equation}
or
\begin{equation}
\Phi_I+\delta\Phi_I = \varphi(r)e^{i\psi_I(r,x,t)}\,,\quad \psi_I(r,t)=k \delta_{Ii} x^i+\delta\psi_I(r)e^{-i\omega t}\,.
\end{equation}
This allows to focus on the dynamics of the phase of the original complex scalars, that is on the phonon dynamics. Plugging the Ansatz $\Phi_I = \varphi(r)e^{i\psi_I(r,x,t)}$ into the complex scalar action \eqref{actionqlattice} and expanding in terms of the fields $\varphi$, $\psi_I$, we restrict our attention to the simplified holographic theory:
\begin{equation}\label{action}
S=\int d^{d+2}x\,\sqrt{-g}\left[R-\frac12\partial\phi^2-\frac14Z(\phi)F^2-V(\phi)-\frac12Y(\phi)\delta^{IJ}\partial\psi_I\partial\psi_J\right],
\end{equation}
which is a generalization of \cite{Andrade:2013gsa}. We have redefined the scalar $\varphi\mapsto\phi(\varphi)$ so that it has a canonically normalized kinetic term. Asymptotically, $\phi(\varphi)\sim\varphi$ but is in general a non-trivial function. The scalar couplings $V$, $Z$ and $Y$ can be related to the couplings in the original action \eqref{actionqlattice} $V_\Phi$, $Z_\Phi$ and $Y_\Phi$. The full background Ansatz is
\begin{equation}
\label{Ansatz}
ds^2=-D(r)dt^2+B(r)dr^2+C(r)d\vec x^2\,,\quad A=A(r)dt\,,\quad \phi=\phi(r)\,,\quad \psi_I=k\delta_{Ii} x^i\,.
\end{equation}
The scalar couplings are arbitrary, we just specify their UV ($\phi\to0$) behavior:
\begin{equation}
\label{UVscalarcouplings}
V_{UV}=-d(d+1)+\frac12m^2\phi^2+\dots,\quad Z_{UV}=1+z_1\phi+\dots,\quad Y_{UV}=y_2 \phi^2+\dots
\end{equation}
which ensures the existence of asymptotically locally AdS$_{d+2}$ black holes geometry when $r\to0$. 
The UV behavior of $Y(\phi)$ is motivated by the complex scalar construction above and is crucial in order to allow for translations to be broken spontaneously. 
Close to the boundary $r\to0$, the scalar $\phi$ behaves as 
\begin{equation}\label{eq:phiBC}
\phi(r\to0)=\phi_{(s)}r^{d+1-\Delta}+\phi_{(v)}r^{\Delta}+\dots\,,\qquad m^2=\Delta(\Delta-d-1)\,.
\end{equation}
By convention, $\Delta>(d+1)/2$ is the largest root of the quadratic polynomial, so $\phi_{(s)}$ is the source (the slowest decaying mode) and $\phi_{(v)}$ the vev (the fastest decaying mode). From our previous discussion, it is clear that when $\phi_{(s)}=0$, translations are broken spontaneously, \cite{Argurio:2014rja,Amoretti:2016bxs}.

When $Y(\phi)=\phi^2$ exactly, the $d+1$ scalars can be combined into $d$ complex scalars $\Phi_I=\phi \exp(i\sqrt2\psi_I)/\sqrt2$,
and the action \eqref{action} can be rewritten as an action for $d$ complex scalar contained within \eqref{actionqlattice}. 
\eqref{UVscalarcouplings} shows this mapping can always be performed asymptotically and so our simplified action \eqref{action} can still be thought of as a CFT deformed by complex operators.\footnote{In general, the $\Phi_I$ can always be defined in terms of a formal integral in the target space over $(\phi,\psi_I)$, but this integral cannot always be evaluated exactly in terms of a simple function.}

We also emphasize that we do not expect the global shift symmetry to be an exact symmetry of the system at all energy scales. 
It represents an emergent low energy symmetry related to the dynamics of the Goldstones. 
Indeed it is absent from the holographic actions where inhomogeneous spatially modulated phases have been studied. 
Nevertheless, since we focus on low energy dynamics in this work, we regard this symmetry as an exact symmetry at all energy scales.

The Goldstone modes can be identified by acting on the background with the Lie derivative along $\partial/\partial_{\vec x}$. It leaves all fields invariant except the $\psi_I$'s. This confirms that phonon dynamics will be captured by the fluctuations $\delta\psi_I$. As we have chosen the same value of $k$ in all spatial directions, the dual state preserves isotropy. Clearly this can be relaxed, with translations spontaneously broken anisotropically along one or several spatial directions.

In the remainder of this work, we set $d=2$.
Assuming the existence of a regular horizon at $r=r_h$, the temperature $T$ and the entropy density $s$ are given by:
\begin{equation}
s=4 \pi C(r_h) \ , \qquad T=\frac{1}{4 \pi }\left.\sqrt{-\frac{B'(r)D'(r)}{B(r)^2}}^{\ '}
\right|_{r=r_h}\ ,
\end{equation}
with the following near-horizon expansion
\begin{equation}
\label{NearHorizon}
\begin{split}
&ds^2=-4\pi T(r_h-r)dt^2+\frac{dr^2}{4\pi T(r_h-r)}+\frac{s}{4\pi}(dx^2+dy^2)+\dots \ ,\\
&A_t=A_h(r_h-r)+\dots\ , \qquad \phi=\phi_h+\dots\ .
\end{split}
\end{equation}

\subsubsection{Holographic renormalization, one-point functions and Ward identities \label{HoloRen}}

In this section we employ holographic renormalization techniques \cite{deHaro:2000vlm} to compute the dual one-point functions and Ward identities.
For simplicity, we set $m^2=-2$. Restricting to the spontaneous case $\phi_{(s)}=0$, the UV expansion of the background in Fefferman-Graham gauge reads:
\begin{equation}
\label{UVexpansionbackground}
\begin{split}
&D(r)=\frac1{r^2}\left(1+d_3 r^3+\mathcal O(r^4)\right)\ , \qquad
B(r)=\frac1{r^2}\ , \qquad
C(r)=\frac1{r^2}\left(1-\frac{d_3}2 r^3+\mathcal O(r^4)\right)\ , \\
&\qquad \qquad \qquad
\phi(r)=\phi_{(v)} r^2+\mathcal O(r^4)\ , \qquad
A(r)=\mu - \rho r+\mathcal O(r^3) \ ,\quad \psi_I=k \delta_{Ii} x^i
\end{split}
\end{equation}
where subleading coefficients are fixed in terms of the vevs $\rho$, $\phi_{(v)}$, $d_3$.

In order to obtain the pressure, we need to compute the background renormalized on-shell action. The 
necessary
boundary counterterms 
are:
\begin{equation}\label{counter}
S_{\text{c.t.}}=\int_{r=\epsilon} d^3x \sqrt{-\gamma} \left[2\mathcal K+4+R[\gamma]+\frac{1}{2}\phi^2-\frac{1}{2}Y(\phi) \sum_{I=1}^2 
\left(\psi_I - k\delta_{Ii} x^i\right)^2\right] \ ,
\end{equation}
where $\gamma_{\mu \nu}$ is the induced metric at $r=\epsilon$ where $\epsilon$ is a UV regulator and $\mathcal K$ is the trace of the extrinsic curvature. As we now explain, the form of the scalar counterterms in \eqref{counter} can be worked out in two ways: i) either in accordance with the symmetries preserved by the background ansatz \eqref{Ansatz}, ii) or by mapping the scalars asymptotically to the complex parameterization (\emph{i.e.} like in the Q-lattice studied for instance in \cite{Amoretti:2016bxs}) .

The background Ansatz \eqref{Ansatz} breaks the original shift and translational symmetries to their diagonal subgroup. The last counterterm in \eqref{counter} does respect this symmetry: the shifts of the $\psi_I$'s are compensated by how $k x^i$ transforms under spatial translations. Holographic renormalization requires that the divergences at first and second order in the fluctuations cancel and that the coefficients of the counterterm do not depend on the asymptotic modes of the background fields. This fixes the form of \eqref{counter} and the values of the numerical coefficients univocally.

As we already explained, due to the UV behavior of the couplings \eqref{UVscalarcouplings}, the model is asymptotically equivalent to the theory of two complex scalar fields $\Phi_I = \phi\, e^{i \psi_I}$, $I=(x,y)$. One can therefore consider the standard counterterm $\Phi_I^* \Phi^I$ needed to renormalize the theory of two massive complex scalars. Specifically, one must first take the variations of $\Phi_I^* \Phi^I$ and then rewrite
 the fluctuations of the complex fields in terms of the fluctuations of the modulus $\phi$ and phases $\psi_I$%
\footnote{Performing these steps in the reverse order leads to a different (wrong) result. In fact, if we express $\Phi_I^* \Phi^I$ in `polar' parametrization (\emph{i.e.} in terms of $\phi$ and $\psi_I$) before considering its variation, we only obtain a counterterm $\phi^2$ which does not renormalize the $\psi_I$ sector.}. This procedure yields scalar counterterms that agree with the last one in \eqref{counter} order by order in those fluctuations.

The renormalized on-shell action for the background is:
\begin{equation}\label{Sbg_ren}
S_{\text{ren}}=\lim_{\epsilon \rightarrow 0}\int_{r=\epsilon} d^3 x \left[-k^2 I_Y(\epsilon)+\sqrt{B(\epsilon) D(\epsilon)}\left(\frac{4 C(\epsilon)}{\sqrt{B(\epsilon)}}+\frac{C'(\epsilon)}{B(\epsilon)}+\frac{C(\epsilon) D'(\epsilon)}{B(\epsilon) D(\epsilon)} \right)+\frac{1}{2}\sqrt{D(\epsilon)}C(\epsilon) \phi(\epsilon)^2 \right] \ ,
\end{equation}
where 
\begin{equation}
\label{IYdef}
 I_Y(r)=\int_{r_h}^r \sqrt{BD}\,Y(\phi) \ .
\end{equation} 
Note that the scalar counterterms in \eqref{counter} do not contribute at the background level, but are necessary to renormalize the action at quadratic order in the fluctuations. 
Evaluating \eqref{Sbg_ren} on the background \eqref{UVexpansionbackground} and continuing through Euclidean signature $t=-i\tau$, $S_{ren}=i I_{ren}$, we obtain for the Euclidean on-shell action:
\begin{equation}
I_{\rm ren}=\beta V_{(2)} \left( -k^2 I_Y(0)+\frac{3 d_3}{2} \right)\ ,
\end{equation}
where $V_{(2)}$ is the boundary spatial volume and $\beta$ the inverse temperature. The pressure is obtained from:
\begin{equation}
\label{pressure}
p=-w=-\frac{I_{ren}}{\beta V_{(2)}}=-\frac{3 d_3}{2}+k^2 I_Y(0) \ .
\end{equation}

In order to compute the energy density we work out the renormalized on-shell action at linear level in the fluctuations. 
We consider the following perturbation of the background fields:
\begin{eqnarray}
g_{\mu \nu}&=&g_{\mu \nu}^b(r)+h_{\mu \nu}(x_{M}) \ ,\\
A_{\mu}&=&A_{\mu}^b(r)+\delta A_{\mu} (x_M) \ , \\
\phi&=&\phi^b(r)+\delta \phi(x_{M}) \ ,\\
\psi_I&=& \psi_I^b(r)+\delta \psi_I(x_{M}) \ ,
\end{eqnarray}
where the fields with the apex $b$ are the background \eqref{Ansatz}, the Greek indices run over the boundary coordinates, and the capital Latin indices run over the whole bulk coordinates (note that we fixed the radial gauge). Using the background EOMs one can easily verify that the action \eqref{action} reduces to a boundary term:
\begin{equation}\label{regularizedlinear}
S_{\rm reg}^{(1)}= \int_{r=\epsilon} d^3x \sqrt{-g^b}\left[ \nabla_{\nu} h^{r \nu}-\nabla^{r} h^{\nu}_{\nu}-\delta \phi \partial^{r} \phi^b-Y(\phi) \sum_{I} \delta \psi_I \partial^{r} \psi_I^b-Z(\phi) \delta A_{\nu} F^{r \nu} \right] \ ,
\end{equation}
where the covariant derivatives and the raising/lowering of indices are both done with the background metric $g_{\mu \nu}^b$. 
The action must be renormalized by adding the counterterms \eqref{counter} expanded to linear order in the fluctuations. 
Focusing on the Ansatz \eqref{Ansatz} and on the asymptotic UV expansion \eqref{UVexpansionbackground}, the fluctuations behave asymptotically as:
\begin{eqnarray}
h_{\mu \nu}&=& \frac{1}{r^2} \left(h^{(0)}_{\mu \nu}(x_{\mu})+h^{(1)}_{\mu \nu}(x_{\mu})r+h^{(2)}_{\mu \nu}(x_{\mu})r^2+h^{(3)}_{\mu \nu}(x_{\mu})r^3+... \right) \ , \\
\delta A_{\mu}&=& \delta A^{(0)}_{\mu}(x_{\mu})+\delta A^{(1)}_{\mu}(x_{\mu}) r+... \ , \\
\delta \phi&=& \delta \phi^{(s)} r +\delta \phi^{(v)} r^2 +... \ , \\
\delta \psi_I&=&\frac{ \delta \psi_I^{(s)}}{r}+\delta \psi_I^{(v)}+... \ .
\end{eqnarray} 
Notice the unusual asymptotic expansion of $\delta\psi_I$, \cite{Argurio:2014rja,Amoretti:2016bxs}. Recalling that in the complex parameterization, $\delta\psi_I(r)=-i\delta\phi(r)/\phi(r)$, we see that it is a direct consequence of $\phi_{(s)}=0$. If $\phi_{(s)}\neq0$, then we would have $\delta \psi_I= \delta \psi_I^{(s)}+\delta \psi_I^{(v)}r^3+... $, as expected for explicit translation breaking \cite{Andrade:2013gsa}.

Then, the renormalized action reads: 
\begin{equation}\label{renspontaneous}
S_{\rm ren}^{(1)}=\int d^3x \left[\frac{3}{2} d_3 h_{tt}^{(0)}+\frac{3 d_3}{4} h_{xx}^{(0)}+\frac{3 d_3}{4} h_{yy}^{(0)}-\rho \delta A_t^{(0)} -\phi_{(v)} \delta \phi^{(s)}\right].
\end{equation}
From \eqref{renspontaneous} one can compute the expectation value of the stress-energy tensor, current and scalars:%
\footnote{Recalling the asymptotic relation between the `polar' scalar fields $(\phi,\psi_I)$ to the complex ones $\Phi_I$, one can see the vanishing of $\langle O_{\psi_I} \rangle$ as a consequence of a non-trivial cancellation between two contributions with opposite signs to the linear on-shell action \eqref{renspontaneous}. This result is in harmony with previous analyses of the Q-lattice \cite{Amoretti:2016bxs}. The correct intuition comes from observing that the Q-lattice can be equivalently thought of as a theory of $O$ or $O^*$.}
 \begin{equation}
\begin{split}\label{stressenergytensor}
\langle T^{tt} \rangle= \epsilon=- 3 d_3 \ , \qquad &
\langle T^{xx} \rangle=\langle T^{yy} \rangle=- \frac32 d_3=p-k^2I_Y(0) \ , \qquad
\langle J^{t} \rangle=\rho \ ,\qquad
\langle O_\phi \rangle=\phi_{(v)} \ , 
\end{split}
\end{equation}
as well as the one-point Ward identities:
\begin{equation}\label{war_ide}
\langle T^{\mu}_{\mu} \rangle=0 \ , \qquad  \partial_{\mu}\langle T^{\mu \nu} \rangle=0 \ ,\qquad  \partial_{\mu}\langle J^{\mu} \rangle=0 \ ,
\end{equation}
where $d_3$ is defined in \eqref{UVexpansionbackground}, the pressure $p$ is given by \eqref{pressure} and $I_Y(0)$ by \eqref{IYdef}. The vanishing of the right hand side in the Ward identities \eqref{war_ide} is consistent with translations being broken spontaneously (though homogeneity of our setup makes this somewhat trivial for the background \cite{Amoretti:2016bxs}).

The equilibrium stress-tensor of an isotropic, conformal crystal is \cite{chaikin_lubensky_1995,Delacretaz:2017zxd}:
\begin{equation}\label{eq:tijcrys}
\langle T^{ij}_{\rm eq}\rangle=
\left[p+\left(G+K\right)\partial\cdot\langle\Psi\rangle\right]\delta^{ij}+2G\left[\partial^{(i}\langle\Psi^{j)}\rangle-\delta^{ij}\partial\cdot\langle\Psi\rangle\right]\ ,
\end{equation}
with $K$ and $G$ the bulk and shear moduli respectively. The bulk modulus only contributes to diagonal elements, the shear modulus only to off-diagonal elements. We find that \eqref{stressenergytensor} is compatible with \eqref{eq:tijcrys} provided we consider a uniform, non-zero strain (phase gradient) $\partial\cdot\langle \Psi\rangle=\bar u$. Then, 
\begin{equation}
\label{eq:phasegradient}
K\bar u=-\frac{k^2}2\int_{r_h}^0 dr\sqrt{BD}Y
\end{equation}
which is positive with our choice of bounds on the integral. This expression is exact in $k$. This state bears some similarity with a superfluid state with a non-zero, uniform superfluid velocity, which also features a non-zero phase gradient. These states have typically a higher free energy than states with no superfluid velocity.  Clearly, from our result \eqref{pressure}, the same is true in our case: the free energy is minimized by setting $k=0$. In this case, there would be no translation breaking left at all. As we will discuss in section \ref{section:HD}, stable phases with $k\neq0$ can be found by including higher-derivative corrections to our original model \eqref{action}.
In \cite{Amoretti:2017axe}, we comment on the potential relevance of these unstable equilibrium states to the strange metals.

We can now boost the stress-energy tensor at rest \eqref{stressenergytensor} with a quadrivelocity $u^{\mu}=(1, \vec{v})$:
\begin{equation}
\label{RelDualST}
\langle T^{\mu \nu}\rangle=\left(p+2K\bar u\right) \eta^{\mu \nu}+\left(\epsilon+p+2K\bar u\right)u^{\mu}u^{\nu},\quad \epsilon=2p+4K\bar u \ .
\end{equation}
The last equation encodes tracelessness of the dual stress-energy tensor due to conformal symmetry of the dual field theory in the UV. It manifestly differs from the equivalent equation for a $2+1$-dimensional conformal fluid or solid without strain, which would read $\epsilon=2p$.

The background EOMs give rise to two radially conserved quantities.
The first simply gives the UV charge density and relates it to the electric flux emitted from the horizon:
\begin{equation}
\label{density}
\langle J^t\rangle = \rho u^t=\rho = \sqrt{-g}Z(\phi)F^{rt}=-\frac{C Z(\phi)}{\sqrt{B D}}A'=\lim_{r\to0}\frac1{r^3}Z(\phi) n_M F^{M t}\,,
\end{equation}
where at equilibrium $u^\mu=(1,0,0)$ and by convention $n_M = (0,1,0,0)$ is the unit (outward-pointing) vector normal to the boundary.
The second radially conserved quantity is defined by the relation:
\begin{equation}\label{radial1}
\left[\rho A(r)+\frac{C^2(r)}{\sqrt{B(r)D(r)}}\left(\frac{D(r)}{C(r)}\right)'-k^2I_Y(r)\right]'=0\,.
\end{equation}
This is the Noether charge associated to the bulk time-like Killing vector \cite{Papadimitriou:2005ii}.
Evaluating \eqref{radial1} both at the horizon and at the boundary using the background asymptotics \eqref{UVexpansionbackground}, we obtain:
\begin{equation}\label{smarr}
s T=- \mu \rho -\frac{9 d_3}{2}+k^2 I_Y(0)\,,
\end{equation}
recovering the usual Smarr law $\epsilon+p= \mu \rho+Ts$.


\subsection{Higher-derivative model: thermodynamically stable phases \label{section:HD}}

\subsubsection{A model for thermodynamically stable phases \label{section:stable}}
As discussed below \eqref{eq:phasegradient}, the two-derivative model \eqref{action} does not allow for classical solutions to the eoms which both have $k\neq0$ and minimize the free energy. This deficiency can be remedied by adding to the action the following higher-derivative terms
\begin{equation}\label{actionHD}
S=\int d^{4}x\,\sqrt{-g}\left[R-\frac12\partial\phi^2-V(\phi)-\frac14\left(Z_1(\phi)+\lambda_1 Z_2(\phi)\sum_{I=1}^{2}\partial\psi_I^2\right)F^2-\frac12\sum_{I=1}^{2}\left(Y_1(\phi)\partial \psi_I^2+\lambda_2 Y_2(\phi)\left(\partial \psi_I^2\right)^2\right)\right].
\end{equation}
Here there is no implicit summation on $I$ indices.
The equations of motion are
\begin{equation}
\begin{split}
0=&G_{\mu\nu}-\frac12\left(Z_1(\phi)+\lambda_1 Z_2(\phi)\sum_{I=1}^{2}\partial\psi_I^2\right)F_{\mu\rho}F_{\nu}{}^\rho-\frac12\nabla_\mu\nabla_\nu\phi\\
&-\frac12\sum_{I=1}^{2}\partial_\mu\psi_I\partial_\nu\psi_I\left(Y_1(\phi)+2\lambda_2Y_2(\phi)\partial \psi_I^2+\frac{\lambda_1}2Z_2(\phi)F^2\right)\\
&+\frac{g_{\mu\nu}}4\left(2V(\phi)+\partial\phi^2+\frac12Z_1(\phi)F^2+\sum_{I=1}^{2}\partial\psi_I^2\left(Y_1(\phi)+\lambda_2Y_2(\phi)\partial\psi_I^2+\frac{\lambda_1}{2}Z_2(\phi)F^2\right)\right),
\end{split}
\end{equation}
\begin{equation}
0=\nabla_\mu\left(\left(Z_1(\phi)+\lambda_1 Z_2(\phi)\sum_{I=1}^{2}\partial\psi_I^2\right)F^{\mu\nu}\right),
\end{equation}
\begin{equation}
0=\Box\phi-\frac14\left(Z_1'(\phi)+\lambda_1  Z_2'(\phi)\sum_{I=1}^{2}\partial\psi_I^2\right)F^2-V'(\phi)-\frac12\sum_{I=1}^{2}\left(Y_1'(\phi)\partial \psi_I^2+\lambda_2Y_2'(\phi)\left(\partial \psi_I^2\right)^2\right),
\end{equation}
\begin{equation}
0=\nabla_\mu\left(\left(Y_1(\phi)+\partial\psi_I^2\left(2\lambda_2Y_2(\phi)+\frac{\lambda_1}2Z_2(\phi)F^2\right)\right)\nabla^\mu\psi_I\right),\quad I=1,2\,.
\end{equation}

Following our previous logic, the extra terms are inspired by expanding extra higher-derivative terms $(\partial\Phi_I\partial\Phi_I^{*})F^2$ and $(\partial\Phi_I\partial\Phi_I^{*})^2$ in a complex scalar action like \eqref{actionqlattice}. In the UV, we assume that $V$, $Z_1$ and $Y_1$ behave as in  \eqref{UVscalarcouplings} while
\begin{equation}
\label{eq:UVscalarcouplingsHD}
Z_2(\phi)\sim z_{2,2}\phi^2+\dots\,,\qquad Y_2(\phi)=y_{2,2}\phi^2+\dots
\end{equation}
We have slightly relaxed the UV behaviour of $Y_2$ compared to the parent complex scalar term, which would dictate $Y_2(\phi)\sim\phi^4$. To the best of our knowledge, this does not affect the holographic renormalization and one-point functions. A coupling $Y_2(\phi)$ as in \eqref{eq:UVscalarcouplingsHD} allows to trigger $k\neq0$ instabilities of the Reissner-Nordström black hole. $Y_2(\phi)\sim\phi^4$ would not allow for instabilities of RN-AdS, though we expect it would lead to instabilities of black holes with $\phi\neq0$. However, they are technically more complicated to exhibit, so we choose \eqref{eq:UVscalarcouplingsHD} for simplicity.

With $\phi=0$, the effects of the term with coupling $\lambda_1$ on the conductivity and charge diffusivity have been considered previously with explicit translation symmetry breaking boundary conditions \cite{Gouteraux:2016wxj,Baggioli:2016pia} (see also \cite{Baggioli:2016oqk}). 
To our knowledge, the $\lambda_2$ coupling has not been explicitly considered in
previous works, but it is implicitly included in models of holographic massive gravity with a general potential for the St\"uckelberg scalars \cite{Baggioli:2014roa,Alberte:2015isw,Alberte:2016xja,Garcia-Garcia:2016hsd}.
 We emphasize though that our setup differs in that the UV boundary conditions are such that translations are spontaneously broken rather than explicitly.
 
Mutatis mutandis, the holographic renormalization of the model proceeds as in section \ref{HoloRen}. We find the same result for the one-point functions:
  \begin{equation}
\label{stressenergytensorHD}
\langle T^{tt} \rangle=- 3 d_3 =2 \langle T^{ii} \rangle \ , \qquad
\langle J^{t} \rangle=\rho \ ,\qquad
\langle O_\phi \rangle=\phi_{(v)} \ .
\end{equation}
However, the free energy now receives extra contributions
 \begin{equation}
 \label{wHD}
w=-p =- \frac{I_{ren}}{\beta V_{(2)}}=\frac32 d_{(3)}-k^2 I_{Y_1}(0)-2\lambda_2 k^4 I_{Y_2}(0)+\lambda_1\rho^2 k^2 I_{Z_2}(0)\,,
\end{equation}
where we have defined
\begin{equation}
\label{IYdefHD}
 I_{Y_1}(r)=\int_{r_h}^r \sqrt{BD}\,Y_1(\phi) \ ,\quad  I_{Y_2}(r)=\int_{r_h}^r \frac{\sqrt{BD}}C\,Y_2(\phi) \ ,\quad  I_{Z_2}(r)=\int_{r_h}^r \frac{Z_2(\phi)A'^2}{\sqrt{BD}}\ .
\end{equation} 
The free energy should be minimized with respect to $k$ to find the most stable phase. This is equivalent to imposing periodic boundary conditions on the spatial coordinates $x^i$, with periodicity $L_x=2\pi/k$. Indeed, this is exactly what we want to describe CDW states. Taking into account isotropy, this leads to $w=-\langle T^{ii}\rangle$ \cite{Donos:2013cka,Donos:2015eew,Donos:2018kkm}. Using \eqref{wHD} and \eqref{stressenergytensorHD}, we get
\begin{equation}
\label{StabilityCond}
 I_{Y_1}(0)+\lambda_2 k^2 I_{Y_2}(0)-\frac12\lambda_1 I_{Z_2}(0)=0\,,
\end{equation}
so that in the end
\begin{equation}
\label{wstable}
w=-p=-\langle T^{ii}\rangle=\frac32 d_3\,.
\end{equation}
The stress-tensor now matches that of an isotropic crystal without an equilibrium phase gradient. As before, we can boost the stress tensor at rest and obtain
\begin{equation}
\label{RelDualSTstable}
\langle T^{\mu \nu}\rangle=p \eta^{\mu \nu}+\left(\epsilon+p\right)u^{\mu}u^{\nu},\quad \epsilon=2p \ ,
\end{equation}
which agrees with similar expressions in \cite{Donos:2013cka,Donos:2015eew,Donos:2018kkm}.

For future reference, we also collect the following expressions
\begin{equation}
\rho=\left.-\frac{C(r)A'(r)}{\sqrt{BD}}\left(Z_1(\phi)+2\lambda_1 k^2\frac{Z_2(\phi)}{C}\right)\right|_{r=r_h}\,,
\end{equation}
\begin{equation}
sT=\left.-\rho A(r)-\frac{C^2(r)}{\sqrt{B(r)D(r)}}\left(\frac{D(r)}{C(r)}\right)'+k^2 I_{Y_1}(r)+2\lambda_2 k^4 I_{Y_2}(r)-\lambda_1 k^2 I_{Z_2}(r)\right|_{r=r_h}\,.
\end{equation}

We now turn to the question of how these phases breaking translations spontaneously arise as the endpoint of instabilities of the (translation invariant) Reissner-Nordström black hole.

\subsubsection{Dynamical instabilities of the Reissner-Nordström black hole \label{subsection:instabilities}}

For simplicity, we require that Reissner-Nordstr\"om is a solution of the equations of motion derived from \eqref{actionHD}. To this end, we consider the following IR expansions of the couplings around $\phi=0$:
\begin{equation}
V(\phi)\sim-6+\frac12m^2\phi^2+\dots\ ,\quad Z_1(\phi)\sim1+ z_2\phi^2+\dots\ ,\quad Z_2(\phi)\sim z_2\phi^2+\dots\ ,\quad Y_1(\phi)=Y_2(\phi)\sim y_2\phi^2+\dots 
\end{equation}
In the IR, it becomes an AdS$_2\times$R$^2$ geometry:
\begin{equation}
\label{ads2DW}
ds^2=-\frac{dt^2}{\xi^2}+\frac{d\xi^2}{6\xi^2}+dx^2+dy^2\,,\quad\phi=0\,,\quad A_t=\frac{\sqrt2}\xi\,,\quad \psi_I=0\,.
\end{equation}
Importantly, we are setting $k=0$ (or equivalently $\psi_I=0$) in the solution \eqref{ads2DW}, as our starting point are translation invariant solutions. 

We now consider radial perturbations of the scalar fields $\phi$, $\psi_I$ around this solution
\begin{equation}
\delta\phi=\phi_0 \xi^{\delta_\phi},\qquad \delta\psi_I=k\delta_{Ii} x^i\ .
\end{equation}
The equations of motion for the $\psi_I$'s are automatically satisfied by our Ansatz. Having $\phi=0$ in the background simplifies our task as the radial perturbations involving the scalar decouple from those of other fields. However, there is no conceptual obstacle to repeating this procedure over an AdS$_2\times$R$^2$ domain-wall with $\phi\neq0$. The main technical obstacle is that perturbations of the scalars do not decouple from other fields, and solving the resulting system of linear equations is somewhat involved.

The IR dimension $\delta_\phi$ of the operator dual to $\phi$ is easily obtained from the equation of motion for $\phi$
\begin{equation}
\label{eq:deltaphi}
\delta_\phi=-\frac12+\frac16\sqrt{9+6m^2-72 z_2+12  k^2\left(y_2-12 z_2\lambda_1\right)+12y_2\lambda_2  k^4}\ .
\end{equation}
Equation \eqref{eq:deltaphi} matches the one in section 3 of \cite{Donos:2014uba} after suitable identifications of the parameters and setting $\lambda_1=\lambda_2=0$.

There is an instability whenever the radicand $\Delta$ changes sign from positive to negative. In order for this instability to be towards a phase with $k\neq0$ (and so breaking translations spontaneously), we need  $\Delta(k)<0$ for $0< k_-< k< k_+$ where $\Delta(k_\pm)=0$.

 It is straightforward to check that this can easily happen in the allowed parameter space on $\lambda_{1,2}$, depending on the specific choice of scalar couplings. The couplings $\lambda_{1,2}$ are constrained by causality: \cite{Gouteraux:2016wxj} found a necessary condition on $\lambda_1$, $-1/6<\lambda_1<1/6$. We take $\lambda_2>0$ and defer a more thorough analysis to future work. These couplings do not result into Shapiro time advances \cite{Camanho:2014apa}, as they do not involve derivatives of the metric.

For concreteness, we consider a model inspired by \cite{Donos:2014uba}:
\begin{equation}
\label{DGmodel}
V(\phi)=-6\cosh\left(\frac{\phi}{\sqrt3}\right)\ , \quad Z_1(\phi)=\cosh^{\gamma/3}\left(\sqrt3\phi\right)\ ,\quad Z_2(\phi)=\frac12\gamma\sinh^2\left(\phi\right)\ , \quad Y_{1,2}(\phi)=12\sinh^2\left(\frac{\phi}{\sqrt3}\right)\ ,
\end{equation}
for which the regime of dynamical instability is
\begin{equation}
\label{AllowedParSpace}
\begin{split}
&\qquad \qquad  \gamma<-4\ ,\quad -\frac16<\lambda_1<\frac{2}{3\gamma}\ ,\quad 0<\lambda_2<-\frac{(2-3\gamma\lambda_1)^2}{(1+12\gamma)}\  ,\\
&\tilde k_-<\tilde k<\tilde k_+\ ,\quad \tilde k_\pm=\frac{1}{2} \sqrt{\frac{3 \gamma  \lambda_1 }{\lambda_2 }-\frac{2}{\lambda_2 }\pm\frac{\sqrt{9 \gamma ^2 \lambda_1^2+12 \gamma  \lambda_2 -12 \gamma  \lambda_1 +\lambda_2+4}}{\lambda_2 }} \ .
\end{split}
\end{equation}
It is interesting to note that the new couplings $\lambda_{1,2}$, even for small values, have changed the range of values of $\gamma$ where the dynamical instability lies (which for $\lambda_{1,2}=0$ is $\gamma>-1/12$ \cite{Donos:2014uba}).

In such a case, we also expect a dynamical instability of the non-zero temperature translation-invariant black hole towards a spatially modulated phase, which can be diagnosed by constructing the corresponding normalizable mode at $k\neq0$, see eg \cite{Donos:2011bh} for a concrete example. The outcome of this computation is a so-called `bell curve' which shows the evolution of the critical temperature below which the condensate forms as a function of $k$. The most stable phase is found for $k_\star$ such that $T_c(k_\star)$ is maximum.

\begin{figure}
\begin{tabular}{cc}
\includegraphics[width=.45\textwidth]{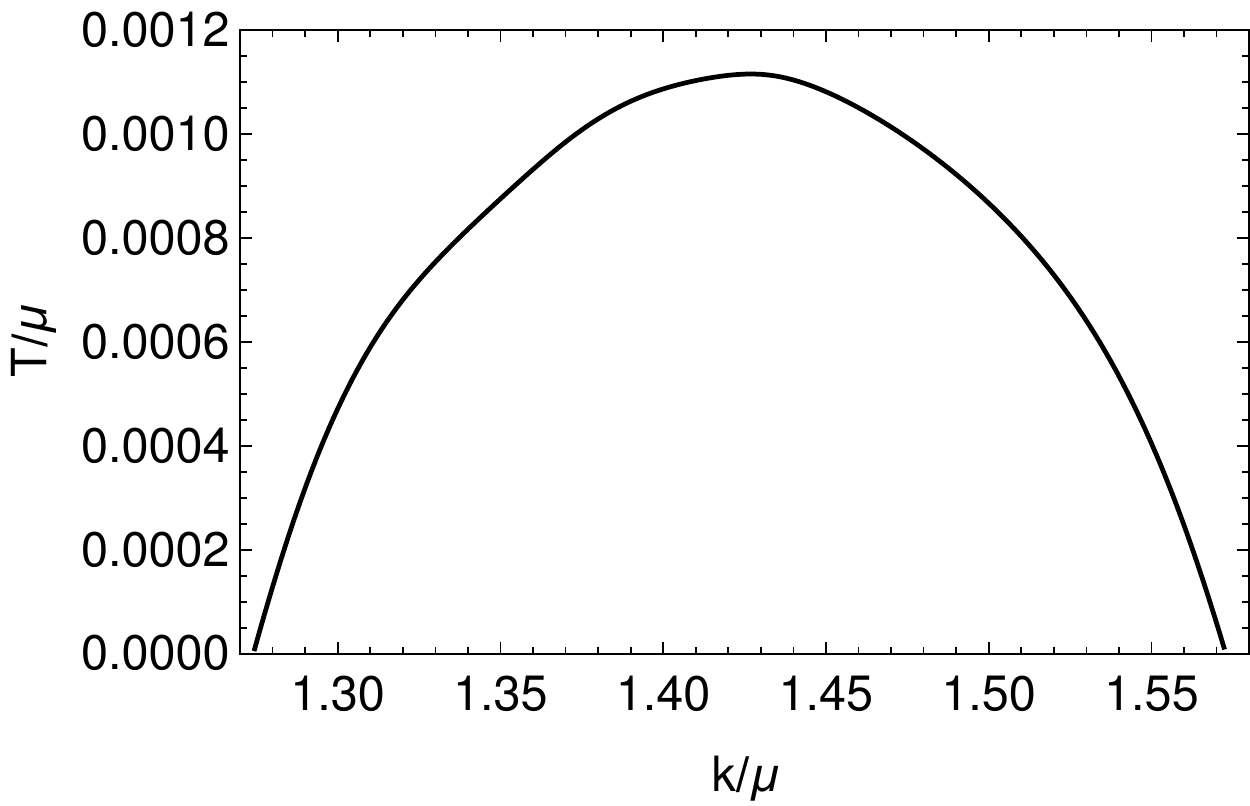}
\end{tabular}
\caption{Instability curve $T_c(k)$ of the Reissner-Nordstr\"om black brane for parameters $\gamma=-6$, $\lambda_1=-0.13$, $\lambda_2=5.10^{-4}$.}
\label{fig:bellcurve}
\end{figure}

We now turn to the construction of such a bell curve in our model \eqref{actionHD} with couplings given by \eqref{DGmodel}.
This implies constructing the unstable mode at non-zero temperature in the Reissner-Nordstr\"om black hole background:
\begin{equation}
\begin{split}
&ds^2=-r^2 f(r)dt^2+\frac{dr^2}{r^2 f(r)}+r^2(dx^2+dy^2)\\
&A_t=\mu\left(1-\frac{r_h}{r}\right)\,,\quad f(r)=1-\frac{r_h^3}{r^3}+\frac{\mu^2r_h}{4r^3}\left(1-\frac{r_h}{r}\right),\quad \phi=\psi_I=0.
\end{split}
\end{equation}
As for zero temperature, the unstable mode obeys a decoupled equation of motion:
\begin{equation}
\delta\phi''+\left(\frac{4}{r}-\frac{f'}f\right)\delta\phi'+\left(-m^2-\frac{2 Y_2 k^2}{r^2}+\frac{2r_h^2\mu^2Z_2\lambda_1 k^2}{r^6}+\frac{-2k^4Y_2\lambda_2+r_h^2\mu^2Z_2}{r^4}\right)\frac{\delta\phi}{r^2f}=0\,.
\end{equation}
We impose regularity at the horizon and spontaneous boundary conditions in the UV. We pick values of $\gamma$ and $\lambda_{1,2}$ satisfying \eqref{AllowedParSpace} and find that this mode exists below a certain critical temperature $T_c(k)$, see figure \ref{fig:bellcurve}. $T_c(k)$ has the bell shape typical in holography. It peaks at a certain critical value $k_\star$, which we expect to be the dynamically preferred value for the backreacted black holes.

\section{The electric conductivity\label{subsection:cond}}

\subsection{Two-derivative model}

When translations are broken spontaneously rather than explicitly, the conductivity carries a pole at $\omega=0$ and takes the general form at low frequencies \cite{Hartnoll:2012rj,RevModPhys.60.1129,Delacretaz:2016ivq,Delacretaz:2017zxd}:
\begin{equation}
\label{ACCond}
\sigma(\omega)\equiv\frac{i}\omega G^R_{JJ}(\omega,q=0)\underset{\omega\to0}{\longrightarrow}\sigma_o+\frac{\chi_{JP}^2}{\chi_{PP}}\frac{i}{\omega}\ .
\end{equation}
$\sigma_o$ is a first-order transport coefficient which appears in the constitutive relation of the current as $J^\mu=\rho u^\mu-T\sigma_o\nabla^\mu(\mu/T)+\dots$ \cite{Delacretaz:2017zxd}, neglecting terms which do not enter in the computation of the conductivity.
$\chi_{JP}$ and $\chi_{PP}$ are the current-momentum and momentum-momentum static susceptibilities. With relativistic symmetry, $\chi_{J P }=\rho$  is simply the charge density.
Similarly, using \eqref{RelDualST}, the momentum susceptibility $\chi_{PP}$ is given by:
\begin{equation}
\label{chiPPdef}
\chi_{PP}\equiv\frac{\delta\langle P^x\rangle}{\delta v_x}=\frac{\delta\langle T^{tx}\rangle}{\delta v_x}= \epsilon+p+2K\bar u=s T + \mu \rho+2K\bar u\ ,
\end{equation}
where in the last step we have used the Smarr relation. Notice the extra contribution compared to the usual expression in relativistic hydrodynamics.

From \eqref{ACCond}, $\sigma_o$ is given by the Kubo formula
\begin{equation}
\label{eq:sigma0Kubo}
\sigma_o=\frac1{\left(\chi_{PP}\right)^2}\lim_{\omega\to0}\frac{i}{\omega}G^R_{J_{\inc}J_{\inc}}(\omega,q=0)\,.
\end{equation}
$J_\inc$ is the incoherent current orthogonal to momentum $P$:
\begin{equation}
\label{JincDef}
J_{\inc} \equiv \chi_{PP} J-\chi_{JP} P \,,\qquad \chi_{j_\inc P}=0\,.
\end{equation}
$\sigma_o$ is an incoherent conductivity, which captures the contribution to \eqref{ACCond} of processes which do not drag momentum.
\cite{Hartnoll:2007ip,Jain:2010ip,Chakrabarti:2010xy,Davison:2015taa} computed $\sigma_o$ analytically for translation-invariant states. It takes a simple form in terms of the background classical solutions to the class of theories \eqref{action} with the fields $\psi_I$ turned off. Here we generalize this computation to the case with spontaneous translation symmetry breaking.

Before we turn to the holographic computation, we must find a set of boundary conditions for which the incoherent current \eqref{JincDef} is sourced but the momentum is not. In order to do this, we note that, by rotating the linear transport relation:
\begin{equation}
\left(\begin{array}{cc}
  J \\
 J_Q \equiv P-\mu J
\end{array} \right)=\left(\begin{array}{cc}
\sigma\; & \alpha T \\
\alpha T\;& \kappa
\end{array} \right)\left(\begin{array}{cc}
\vec{E} \\
 -\frac{\vec{\nabla}T}{T}
\end{array} \right)
\end{equation}
by the matrix:
\begin{equation}
M=\left(\begin{array}{cc}
\chi_{PP}-\mu \rho\; & -\rho \\
\mu\;& 1
\end{array} \right)
\end{equation}
we obtain:
\begin{equation}
\left(\begin{array}{cc}
  J_{\inc} \\
 P
\end{array} \right)=M\left(\begin{array}{cc}
\sigma\; & \alpha T \\
\alpha T\;& \kappa
\end{array} \right)M^T \left(M^T\right)^{-1}\left(\begin{array}{cc}
\vec{E} \\
 -\frac{\vec{\nabla}T}{T}
\end{array} \right) \ .
\end{equation}
This shows that we should impose a set of boundary conditions for which
\begin{equation}
\label{eq:rotationSources}
\left(M^T\right)^{-1} \left(\begin{array}{cc}
\vec{E} \\
 -\frac{\vec{\nabla}T}{T}
\end{array} \right) = \left(\begin{array}{cc}
\bar{E} \\
 0 \end{array}\right) \ ,
\end{equation}
where $\bar{E}$ is the source for the incoherent current. 

We now proceed with the holographic computation. Taking inspiration from \cite{Davison:2015taa,Davison:2017}, we turn on the following set of boundary conditions:
\begin{equation}
\label{pert1}
\delta a_x(r)= a(r)-p_1(r) t\ ,\quad \delta h_{tx}= h_1(r)-p_2(r) t\ ,\quad \delta h_{rx}= h_2(r)\ ,\quad \delta\psi_x=\chi(r)\ .
\end{equation}
Note that we could in principle add a $\delta \psi_0 t$ term in the fluctuation for $\psi_x$, since in this setup this is a vev and does not introduce a new source in the boundary. This term is precisely what acting on the background with the Lie derivative along $x$ would generate, and which we previously identified as the bulk dual to the boundary phonon.
However, we find that it does not contribute to $\sigma_o$ and so do not turn it on to avoid cluttering our expressions. This seems consistent with the intuition that $\sigma_o$ captures the contribution of processes which do not drag momentum, and so should also be insensitive to phonon dynamics.

The $t$ dependence drops out from the linearized equations, provided
\begin{equation}
\label{pert2}
p_1(r)=p_1^{(0)}+\bar{E}\rho A\ ,\qquad p_2(r)=-\bar{E}\rho D \ ,
\end{equation}
where $p_1^{(0)}$ is a constant which will be fixed shortly.

We can now show that the following UV boundary conditions are consistent
\begin{equation}
\label{pert3}
a(r)=a^{(1)}+\mathcal O(r^2)\ ,\quad h_1(r)=h_1^{(1)} r+\mathcal O(r^2)\ ,\quad h_{rx}=\mathcal O(r)\ ,\quad \chi(r)=\chi_1 +\mathcal O(r^3)\ ,
\end{equation}
provided we set 
\begin{equation}
\label{pert4}
p_1^{(0)}=-\bar{E}\left(\frac92 d_3+\rho\mu\right)=\bar E\left(sT-k^2 I_Y(0)\right).
\end{equation}
This condition follows from requesting $\delta h_{rx}$ to fall off sufficiently fast in the UV and is the key difference with the computations in \cite{Donos:2014uba,Donos:2014cya}, which hold for explicit symmetry breaking.

The boundary sources are \cite{Hartnoll:2009sz,Herzog:2009xv}
\begin{equation}
E_x=-\lim_{r\to 0}\partial_t \left(a_x(r,t)+\mu h_t^x(r,t)\right)\ ,\qquad \frac1T\nabla_x T=\lim_{r\to 0}\partial_t h_{t}^x(r,t)
\end{equation}
or, plugging in our boundary expansions 
\begin{equation}
\label{UVsourcesfluctuations}
E_x=\bar{E} \left( s T-k^2 I_Y(0) \right)\,,\qquad \nabla_x T=\bar{E}\rho T
\end{equation}
which verify \eqref{eq:rotationSources} as expected.

We now need to find a radially conserved quantity which asymptotes to $J_\inc$. The $x$ component of Maxwell's equation reads
\begin{equation}
\label{maxeqx}
\left(Z\frac{\sqrt D}{\sqrt B}a'-\frac\rho{C}h_1\right)'=0\,,
\end{equation}
while the $tx$ component of the Einstein equations is
\begin{equation}
\label{eeqtx}
\left(\frac{C^2}{\sqrt B \sqrt D}\left(\frac{h_1}{C}\right)'-\rho a\right)'-k^2\frac{\sqrt B}{\sqrt D}Y h_1=0\,.
\end{equation}
Taking our cue from the definition of $J_\inc$ \eqref{JincDef}, we identify 
$J_\inc(r)=\left(sT-k^2 I_Y(0)\right)$ \eqref{maxeqx} $-\rho D/C$ \eqref{eeqtx}, which is obviously radially conserved $J_\inc'(r)=0$. Explicitly,
\begin{equation}\label{Jinc}
\begin{split}
J_\inc(r)=&-\left[\frac{C^2(D/C)'}{\sqrt D\sqrt B}+k^2\left(I_Y(0)- I_Y(r)\right)\right]\frac{\sqrt D}{\sqrt B}Z a'\qquad \\
&+\rho C\frac{\sqrt D}{\sqrt B}(h_1/C)'+\rho k^2 \left(I_Y(0)- I_Y(r)\right)\frac{h_1}{C}
\end{split}
\end{equation}
Let us now check that this does asymptote to the correct $J_{\inc}$.

At the boundary, the fluctuations have the following behavior:
\begin{eqnarray}
h_1(r)&=&h_1^{(1)}r+\frac{\rho a^{(1)}}{4}r^2+\mathcal{O}(r^3)\ ,\\
h_2(r)&=&h_2^{(1)}r+h_2^{(2)}r^2+\mathcal{O}(r^3)\ ,\\
a(r)&=&a^{(1)}r+\mathcal{O}(r^3)\ .
\end{eqnarray}
The electric current is defined by 
\begin{equation}
J^x=\sqrt{-g}Z(\phi) F^{rx}=\lim_{r\to0}\frac1{r}Z(\phi)n_M F^{M x}=a^{(1)}\ .
\end{equation}
To define the heat current, it is useful to consider the following antisymmetric 2-form:
\begin{equation}
G_{MN}=2\nabla_{M}k_N+ Z(\phi)k_M F_N{}^{S} A_S+\frac12 Z(\phi)(\Xi-2\Theta)F_{MN}
\end{equation}
where $k^M=(1-\rho \bar{E} x,0,0,0)$ is the time-like Killing vector and $\Psi,\Theta$ are defined from
\begin{equation}
\mathcal L_k A=d\Xi\ ,\qquad k^M F_{MN}=\nabla_N\Theta
\end{equation}
such that
\begin{equation}
\Xi=-p_1{}^{(0)} x\ ,\qquad \Theta=-A-p_1 x\,.
\end{equation}
This 2-form obeys the following equation on-shell \cite{Donos:2014cya}
\begin{equation}
\nabla_M G^{MN}=-V(\phi)k^N\,.
\end{equation}
 which follows from the conservation of the bulk Noether charge associated to the time-like Killing symmetry \cite{Papadimitriou:2005ii}.
Projecting this equation on $N=t$ and reabsorbing the right-hand side inside the radial derivative, we recover the radially conserved quantity \eqref{radial1} \cite{Davison:2017}. 
This leads us to identifying the heat vector as $J_Q{}^M=-G^{rM}$. Projecting on $M=t$ and keeping in mind the sign convention $u^t=1$, $J_Q{}^t=s T u^t=s T$. 
Also the heat current reads
\begin{equation}
J_Q{}^x=-G^{rx}=-3 h_1^{(1)}-\mu a^{(1)} .
\end{equation}
Finally, we evaluate the incoherent current \eqref{Jinc} asymptotically and find:
\begin{equation}
J_{\inc}=-\frac92d_3 a^{(1)}+3\rho h_1^{(1)}=\left(s T+2K\bar u\right) J-\rho J_Q=\chi_{PP} J-\rho P
\end{equation}
as it should be. The middle equality shows that $\chi_{P J_Q}=s T+2K\bar u$ in the unstable case, different from the translation-invariant case where $k=0$.

We can now evaluate \eqref{Jinc} on the horizon, imposing regularity of the perturbations in Eddington-Finkelstein coordinates \cite{Donos:2014cya}:
\begin{equation}
\begin{split}
&a=-\bar{E}\frac{s T+2K\bar u}{4\pi T}\log(r_h-r)(1+\mathcal O(r_h-r))\ , \\
&h_1=\frac{2\rho\bar{E} K\bar u}{k^2 Y}+\mathcal O(r_h-r)\ , \\
&h_2=\frac{h_1}{4\pi T(r_h-r)}+\mathcal O(1)\ .
\end{split}
\end{equation}
Plugging this into $J_\inc(r_h)$ and dividing by the source $\bar{E}$, we get the zero frequency limit of the retarded Green's function of the incoherent current and thus, from \eqref{eq:sigma0Kubo} we get:
\begin{equation}
\label{eq:sigmarealpart}
\sigma_o=\frac{\left(s T+2K\bar u\right)^2Z_h}{\left(s T+\mu\rho+2K\bar u\right)^2} +\frac{ 16\pi (K\bar u)^2 \rho^2}{s\,  Y_h k^2\left(s T+\mu\rho+2K\bar u\right)^2}.
\end{equation}
Recalling that $K\bar u\sim k^2$, the $k\to0$ limit is smooth and matches previous results \cite{Hartnoll:2007ip,Jain:2010ip,Chakrabarti:2010xy,Davison:2015taa}. An important difference when $k\neq0$ is that now $\sigma_o$ is not solely expressed in terms of horizon data, but also involves an integral over the whole spacetime. 

We will see that both \eqref{ACCond} together with \eqref{eq:sigmarealpart} and \eqref{chiPPdef} perfectly match our numerical results in section \ref{section:BHs}. 

\subsection{Higher-derivative model}

It is now straightforward to repeat this calculation in the higher-derivative model \eqref{actionHD}. There are two main differences. Firstly, from \eqref{RelDualSTstable}, we find that the momentum static susceptibility now reads
\begin{equation}
\label{chiPPstable}
\chi_{PP}=\frac{\delta\langle  T^{ti}\rangle}{\delta v^i}=-\frac92 d_3=\epsilon+p=sT+\mu\rho\,.
\end{equation}
 once the free energy is minimized with respect to $k$. So the incoherent current will be
\begin{equation}
J_{inc}=\chi_{PP} J-\rho P=sT J-\rho J_Q.
\end{equation}
Secondly, the expressions for the radially conserved bulk currents have to be updated:
\begin{equation}
J^x(r)=\sqrt{\frac{D}B}\left(Z_1+\lambda_1 k^2 \frac{Z_2}{C}\right)a'-\rho\frac{h_1}{C}\ ,
\end{equation}
\begin{equation}
G_{MN}=2\nabla_{M}k_N+ \left(Z_1(\phi)+\lambda_1 Z_2(\phi)\sum_{I=1}^{2}\partial\psi_I^2\right)k_M F_N{}^{S} A_S+\frac12 \left(Z_1(\phi)+\lambda_1 Z_2(\phi)\sum_{I=1}^{2}\partial\psi_I^2\right)(\Xi-2\Theta)F_{MN}
\end{equation}
which leads to the heat current
\begin{equation}
J_Q^x(r)=-G^{rx}=\frac{AA'}{\rho B C}a'+\left(\frac{\rho A}{C}+\frac{D'}{\sqrt{BD}}\right)h_1-\sqrt{\frac{D}B}h_1'\ .
\end{equation}
The incoherent combination of bulk currents
\begin{equation}
\label{JincHD}
J_{inc}(r)=sT J^x(r)-\rho J_Q^x(r)
\end{equation}
is manifestly radially conserved and asymptotes to the incoherent current $J_{inc}=sT J^x-\rho J_Q^x$ at the boundary. 

We can now evaluate \eqref{JincHD} on the horizon, imposing regularity of the perturbations in Eddington-Finkelstein coordinates \cite{Donos:2014cya}:
\begin{equation}
\begin{split}
&a=-\bar{E}\frac{s }{4\pi }\log(r_h-r)(1+\mathcal O(r_h-r))\ , \\
&h_1=\mathcal O(r_h-r)\ , \\
&h_2=\mathcal O((r_h-r)^0)\ .
\end{split}
\end{equation}
we find that 
\begin{equation}
J_{inc}(r_h)=\bar E (sT)^2\left(Z_{1,h}+\lambda_1 k^2\frac{4\pi Z_{2,h}}{s}\right)
\end{equation}
which leads to 
\begin{equation}
\label{sigmaincHD}
\sigma_{inc}=\frac{J_{inc}}{\bar E}=(sT)^2\left(Z_{1,h}+\lambda_1 k^2\frac{4\pi Z_{2,h}}{s}\right)
\end{equation}
and
\begin{equation}
\label{eq:sigmarealpartHD}
\sigma_o=\frac{\sigma_{inc}}{(\chi_{PP})^2}=\frac{\sigma_{inc}}{(sT+\mu\rho)^2}\ .
\end{equation}
This matches the results of \cite{Donos:2018kkm}. Remarkably \eqref{sigmaincHD} has exactly the same functional dependence on horizon data as when translations are explicitly broken (of course the dependence on boundary data $T$, $\mu$ and $k$ will differ since the states are different).

\cite{Delacretaz:2016ivq} pointed out that once translations are weakly broken, the dc conductivity of weakly-disordered, non-Galilean invariant CDWs is precisely given by $\sigma_o$ above, computed directly in the clean theory: this is the physical content of our formula \eqref{eq:sigmarealpartHD}.

\section{Quantum critical CDW phases  \label{section:QCPs}}

\subsection{Two-derivative model \label{section:QCP2D}}

$T=0$ solutions solving the equations of motion deriving from the action \eqref{action}
and modeling holographic quantum critical phases
 were thoroughly studied in \cite{Gouteraux:2014hca} (see also \cite{Donos:2014uba}). The analysis only relies on assuming the following IR behavior $\phi\to\infty$ for the scalar couplings:
\begin{equation}
\label{IRscalarcouplings}
V_{IR}=V_0 e^{-\delta\phi}\ ,\quad Z_{IR}=Z_0 e^{\gamma\phi}\ ,\quad Y_{IR}=Y_0 e^{\nu\phi}\ ,
\end{equation}
and, in principle, is valid irrespectively of the UV boundary conditions.\footnote{Of course, a UV completion is necessary to actually realize these phases.} It carries through in our setup, and allows us to describe quantum critical phases with spontaneous translation symmetry breaking.

The leading order behavior of the fields in the IR is
\begin{equation}\label{skaska}
\begin{split}
& ds^2 = \xi^\theta \left[ -f(\xi)\frac{dt^2}{\xi^{2z}} + \frac{L^2 d\xi^2}{\xi^2f(\xi) } + \frac{d\vec{x}^2}{\xi^2}\right], \quad 
 A = A_0\, \xi^{\zeta - z} dt\ ,\quad
\psi_I = k\delta_{Ii} x^i\,,\quad \phi=\kappa\log \xi\,,\quad f(\xi)=1-\left(\frac{\xi}{\xi_h}\right)^{2+z-\theta}.
 \end{split}
\end{equation}
$\xi_h$ is the location of a Killing event horizon, with the associated Hawking temperature $T\sim \xi_h^{-1/z}$ and Hawking-Bekenstein entropy $s\sim \xi_h^{\theta-2}$. Combining both formul\ae\ the entropy density scales as $s \sim T^{\frac{2-\theta}{z}}$. 

There are four classes of solutions, which differ by whether the fields $\psi_I$ and $A_t$ are related to a marginal or irrelevant deformation of the IR solution. At the level of equations, plugging in a solution of the form \eqref{skaska} returns a system of equations with terms depending on powers of $\xi$. The couplings \eqref{IRscalarcouplings} are either such that all terms in the eoms depend on the same power of $\xi$, and then the eoms reduce to algebraic equations. Or terms involving the Maxwell or $\psi_I$ fields scale with a subleading power of $\xi$ compared to other terms. In this case, they parameterize irrelevant deformations of the leading IR solution, which is then obtained by setting the irrelevant terms to zero in the eoms and solving them. The full IR solution is now a series expansion, where subleading terms are obtained order by order by backreacting the irrelevant terms in the eoms on the leading order solution.

The generic consistency conditions (valid for all classes) are:
\begin{itemize}
 \item Null Energy Condition:
\begin{equation}
\label{NEC}
(2-\theta)(2z-2-\theta)\geq0\ ,\qquad (z-1)(2+z-\theta)\geq0\ .
\end{equation}
 \item Positivity of the specific heat:
 \begin{equation}
 \label{ThermStab}
\frac{2-\theta}{z}\geq0\ .
\end{equation}
 \item Marginal or irrelevant deformation sourced by the  $\psi_I$:
 \begin{equation}
 \label{IrrDefPsi}
-\frac{2+\kappa\nu}{z}\geq0\ .
\end{equation}
This corresponds to an IR operator of dimension $\tilde\Delta_{\psi}=2+z-\theta-\frac{\kappa\nu}2$. The tilde is to emphasize that this is an IR dimension, not the UV engineering dimension of the field $\psi_I$. Still we denote it by $\tilde\Delta_{\psi}$ as the source of this IR operator is proportional (but not equal) to $k$. The operator is marginal when $\kappa\nu=-2$. When it is irrelevant, it backreacts on the leading solution as
\begin{equation}
\label{Irrpsisol}
\Sigma=\Sigma_{k=0}\left(1+c_\Sigma k^2 \xi^{2+\kappa\nu}+O(k^4 \xi^{2(2+\kappa\nu)})\right)
\end{equation}
where $\Sigma$ is a placeholder for the metric, gauge field or scalar $\phi$ and $c_\Sigma$ is a coefficient whose precise form is not important for our discussion. The leading order solution $\Sigma_{k=0}$ is obtained by solving the eoms with the Ansatz \eqref{skaska} setting $k=0$. We see clearly that in the marginal limit $\kappa\nu=-2$ the operator does not source any additional $\xi$ dependence.
 \item Marginally relevant or irrelevant deformation sourced by $A_t$:
 \begin{equation}
 \label{IrrDefAt}
-\frac{\zeta-\theta+2}{z}\geq0\ .
\end{equation}
This corresponds to an IR operator of dimension $\tilde\Delta_{A}=z+1-(\zeta+\theta)/2$. It is marginal when $\zeta=\theta-2$. Similarly to above, when it is irrelevant, it backreacts on the leading solution as
\begin{equation}
\label{IrrAtsol}
\Sigma=\Sigma_{A_0=0}\left(1+c_\Sigma A_0^2 \xi^{\zeta-\theta+2}+O(A_0^4 \xi^{2(\zeta-\theta+2)})\right)
\end{equation}
 The leading order solution $\Sigma_{A_0=0}$ is obtained by solving the eoms with the Ansatz \eqref{skaska} setting $A_0=0$. We see clearly that in the marginal limit $\zeta=\theta-2$ the operator does not source any additional $\xi$ dependence.
\end{itemize}
The dimensions of the IR operators obey:
\begin{equation}
\label{IrrDefCond}
\tilde\Delta_{\psi}\geq3+z-\theta\,,\quad\tilde\Delta_{A}\geq2+z-\theta\,,
\end{equation}
as expected for irrelevant (marginal) operators.
The shift in the condition on $\tilde\Delta_\psi$ originates from the spatial dependence of the source of the IR operator.
Only the sources of these IR operators can be turned on, as the vev term would spoil the IR asymptotics \cite{Davison:2017}.

Next, we discuss the low temperature asymptotics of the incoherent conductivity \eqref{eq:sigmarealpart}. The integral $I_Y(0)\neq0$ is dominated by the UV of the geometry at $T=0$. This can be seen by plugging in the IR geometry \eqref{skaska} and observing that the integrand vanishes in the IR limit. This means that $I_Y(0)$ is going to some constant at $T=0$ which is expressed in terms of UV data and cannot be evaluated solely by the knowledge of the near-horizon region. This is generally the case of static susceptibilities in holography, except for a few special cases \cite{Blake:2016wvh,Davison:2017}. The incoherent conductivity \eqref{eq:sigmarealpart} becomes in the low temperature limit
\begin{equation}
\label{eq:sigmainclowT}
\sigma_o(T\to0)=\frac{ 4(K\bar u)^2}{\left(\mu\rho+2K\bar u\right)^2} \left(Z_h+\frac{ 4\pi \rho^2 }{s k^2 \, Y_h}\right),
\end{equation}
where we have neglected the $s T$ terms which are subleading compared to $\mu\rho$ or $K$ at $T=0$.
It is interesting to note that the expression inside the parentheses is precisely the dc conductivity that would follow from explicit translation symmetry breaking boundary conditions \cite{Donos:2014uba,Gouteraux:2014hca}. Consequently, since the prefactor approaches a constant at $T \rightarrow 0$, the low T dependence of $\sigma_o$ is still completely governed by the near-horizon region. Plugging in the scaling solutions \eqref{skaska}, we find
\begin{equation}
\label{eq:sigmalowTscaling}
\sigma_o\sim\frac{ 4(K\bar u)^2}{\left(\mu\rho+2K\bar u\right)^2} T^{\frac{2z-\theta-2\tilde\Delta}{z}},
\end{equation}
where $\tilde\Delta=max\left(\Delta_{\tilde\rho},\Delta_{\tilde\psi}-1\right)$.
This is the same temperature scaling that was determined in \cite{Gouteraux:2014hca}, assuming explicit translation symmetry breaking.

We conclude this section by taking the `semi-locally critical' limit where $z\to+\infty$, $\theta\to-\infty$ with $\tilde\theta=-\theta/z$ fixed. The scaling \eqref{eq:sigmalowTscaling} now becomes
\begin{equation}
\label{eq:sigmalowTscalingSLC}
\sigma_o\sim \frac{ 4(K\bar u)^2}{\left(\mu\rho+2K\bar u\right)^2} T^{-\tilde\theta}.
\end{equation}
We note two particularly interesting cases: $\tilde\theta=0$ and $\tilde\theta=1$. When $\tilde\theta=0$, the scalar is just a constant in the IR, the entropy and the incoherent conductivity of the state are finite at zero temperature (ie it is AdS$_2\times$R$^2$). When $\tilde\theta=1$, the entropy and $1/\sigma_o$ are both linear in temperature for low temperatures.

\subsection{Higher-derivative model \label{section:QCPHD}}

Scaling solutions of the higher-derivative model can be analyzed along the same lines as in section \ref{section:QCP2D}. We assume the following behaviour of the scalar couplings as $\phi\to\infty$
\begin{equation}
\label{IRscalarcouplingsHD}
V_{IR}=V_0 e^{-\delta\phi}\ ,\quad Z_{IR,i}=Z_{i,0} e^{\gamma_i\phi}\ ,\quad Y_{IR,i}=Y_{i,0} e^{\nu_i\phi}\ ,\quad i=1,2\,.
\end{equation}
Then we look for solutions of the form \eqref{skaska}. We need to decide whether the higher derivative terms parameterize marginal or irrelevant deformations of the leading IR solution. The number of classes is combinatorially larger, and we leave a full analysis of all classes to future work. It is enough for now to comment on a few special cases.

First, we address the case when all terms in the eoms scale with the same power of $\xi$: in this case there are only marginal deformations. Then the solution reads
\begin{equation}
\begin{split}
&\gamma_1=-\nu_2=-\delta-2\nu_1\,,\quad\kappa\nu_1=-\kappa\gamma_2=-2\,,\quad \kappa\delta=\theta\,,\quad\kappa=\sqrt{(\theta-2)(2-2z+\theta)}\,,\quad\zeta=\theta-2\,,\\
& L^2=-\frac{-2 \lambda _1 \lambda _2 k^6 Y_{2,0} Z_{2,0}+2 V_0 \left(\lambda _1 k^2 Z_{2,0}+Z_{1,0}\right)+k^2 Y_{1,0} Z_{1,0}}{2 (-\theta +z+2) \left(\lambda _1 k^2 Z_{2,0} (2 z-\theta )+Z_{1,0} (-\theta +z+1)\right)}\,,\\
&A_0^2=\frac{2 k^2 \left(2 \lambda _2 k^2 Y_{2,0} (-\theta +z+1)+Y_{1,0} (2 z-\theta )\right)+4 V_0 (z-1)}{(-\theta +z+2) \left(-2 \lambda _1 \lambda _2 k^6 Y_{2,0} Z_{2,0}+2 V_0 \left(\lambda _1 k^2 Z_{2,0}+Z_{1,0}\right)+k^2 Y_{1,0} Z_{1,0}\right)}\,.
 \end{split}
\end{equation}
Plugging this in the formula for the incoherent conductivity \eqref{sigmaincHD} returns
\begin{equation}
\label{sigmaincHDmarginal}
\sigma_{inc}=16 \pi ^2 T^2 \xi_h^{\theta } \left(Z_{1,0}+2 \lambda _1 k^2 Z_{2,0}\right)\sim\left(Z_{1,0}+2 \lambda _1 k^2 Z_{2,0}\right) T^{2-\frac\theta{z}}.
\end{equation}
What this result shows is that the low temperature dependence of the incoherent conductivity is set by the two-derivative coupling $Z_1$ of the scalar $\phi$ to the gauge field. The higher derivative coupling $\lambda_1 Z_2$ modifies the prefactor, but not the temperature dependence. It will not cancel out the leading two-derivative term, except in extremely fined tuned circumstances where the prefactor $Z_{1,0}+2 \lambda _1 k^2 Z_{2,0}$ happens to vanish. This is a generic phenomenon: extra higher derivative terms in the EFT sourcing marginal deformations in the IR will act in a similar way.

Let us now address the case where terms coming from $Y_1$ are irrelevant compared to other terms. These terms will act on the leading solution with $k=0$ as in \eqref{Irrpsisol}. We can readily see how this affects the entropy density:
\begin{equation}
\label{sY1irr}
s=4\pi\xi_h^{\theta-2}\left(1+c_s k^2\xi_h^{2+\kappa\nu}+O\left(k^4\xi_h^{2(2+\kappa\nu)}\right)\right)\sim T^{\frac{2-\theta}z}\left(1+c_s k^2T^{-\frac{2+\kappa\nu}z}+O\left(k^4T^{-2\frac{2+\kappa\nu}z}\right)\right).
\end{equation}
$Z_{1,h}$ and $Z_{2,h}$ are affected in a similar way through their dependence on $\phi$.
By the constraint \eqref{IrrDefPsi}, the subleading term vanishes faster than the leading term at low temperatures. It is natural to request that terms coming from $Y_2$ cannot vanish slower in the IR than terms coming from $Y_1$. Otherwise, this would contradict EFT principles. Thus, it is enough to comment on the behaviour of the $Y_1$ terms. Putting together \eqref{sY1irr} and \eqref{sigmaincHD}, we find again that $\sigma_{inc}\sim T^{2-\frac\theta{z}}$.

A more interesting case is when terms coming from $Z_1$ are irrelevant. This can be anticipated, since whether these terms are marginal or irrelevant affects the leading order scaling of $A_t$ through the value of the exponent $\zeta$. It is natural to expect this should have an important consequence on the conductivity. In this case, as in \eqref{IrrAtsol}, the leading order solution is found by setting $A_t=0$ in the eoms and solving them:
\begin{equation}
\label{LeadingSolAtsolZirr}
\begin{split}
&\nu_2=\delta+2\nu_1\,,\quad\kappa\nu_1=-2\,,\quad \kappa\delta=\theta\,,\quad\kappa=\sqrt{(\theta-2)(2-2z+\theta)}\,,\\
& L^2=\frac{\lambda _2 k^4 Y_{2,0}+k^2 Y_{1,0}+V_0}{(\theta -2) (-\theta +z+2)}\,,\quad 2 \lambda _2 k^4 Y_{2,0} (-\theta +z+1)+ k^2Y_{1,0} (2 z-\theta )+2 V_0 (z-1)=0\,.
 \end{split}
\end{equation}
The last equation does not mean that $k$ is fixed in the IR solution. There is a scaling symmetry $x\to L_x x$ which is actually necessary to connect to an asymptotically AdS$_4$ spacetime. 
The leading behaviour of $A_t$ is found by plugging in the $A_t=0$ solution in the $t$ component of Maxwell equation, and solving it for $A_t$. For now we assume that $Z_2$ terms scale the same as $Z_1$ terms. This returns
\begin{equation}
\label{AtsolZirr}
A_t=A_0\xi^{\zeta-z}\,,\quad \kappa\gamma_1=2-\zeta,\,,\quad \kappa\gamma_2=\theta-\zeta\,.
 \end{equation} 
Plugging into \eqref{sigmaincHD}, we find 
\begin{equation}
\label{sigmaincHDirr}
\sigma_{inc}\sim T^{2+\frac{\zeta+2(1-\theta)}{z}}\left(1+O\left(T^{-\frac{\zeta -\theta +2}{z}}\right)\right),
\end{equation}
where the subleading terms come from backreacting \eqref{AtsolZirr} on \eqref{LeadingSolAtsolZirr} and indeed vanish faster than the leading ones at low temperature given \eqref{IrrDefAt}. The same arguments will apply if $Z_2$ terms are subleading compared to $Z_1$ terms (then the condition on $\gamma_2$ in \eqref{AtsolZirr} is relaxed). Notice that setting $\zeta=\theta-2$ recovers the previous scaling \eqref{sigmaincHDmarginal} we obtained in the case of a marginal deformation.

As we elaborate further on in \cite{Amoretti:2017axe}, we can use this result to predict the scaling of the dc conductivity of the same phase weakly pinned by disorder. Indeed, \cite{Delacretaz:2016ivq,Delacretaz:2017zxd} found that the dc conductivity of a pinned CDW is $\sigma_o$ to leading order in the disorder strength, and so can be evaluated in the clean theory. Since $\chi_{PP}$ asymptotes to a constant at zero temperature, its low temperature behavior will be the same as in \eqref{sigmaincHDirr}
\begin{equation}
\sigma_o\sim T^{2+\frac{\zeta+2(1-\theta)}{z}}\,.
\end{equation}
This is one of the main results of our work.
Taking into account the various constraints on the exponents mentioned in section \ref{section:QCP2D}, this can vanish or diverge at low temperatures, ie the CDWs can be either insulating  $\left.d\sigma_o/dT>0\right|_{T\to0}$ or conducting  $\left.d\sigma_o/dT<0\right|_{T\to0}$.

\section{Homogeneous black holes with non-zero strain \label{section:BHs}}

In this section, we construct numerically some examples of homogeneous, finite temperature black holes dual to phases spontaneously breaking translations and with non-zero strain. For this, we restrict to the two-derivative model \eqref{action}. For illustrative purposes, we consider two models, one which contains black holes where the entropy density does not vanish at $T=0$, and another where it does. The first is simply \eqref{DGmodel} with $\gamma=\{1,4\}$. The second is
\begin{equation}
\label{model2}
V(\phi)=-6\cosh\left(\frac{\phi}{\sqrt3}\right)\,,\quad Z(\phi)=\exp\left(-\sqrt3\phi\right)\,,\quad Y(\phi)=\left(1-\exp\phi\right)^2\,.
\end{equation}

\begin{figure}
\begin{tabular}{cc}
\includegraphics[width=.45\textwidth]{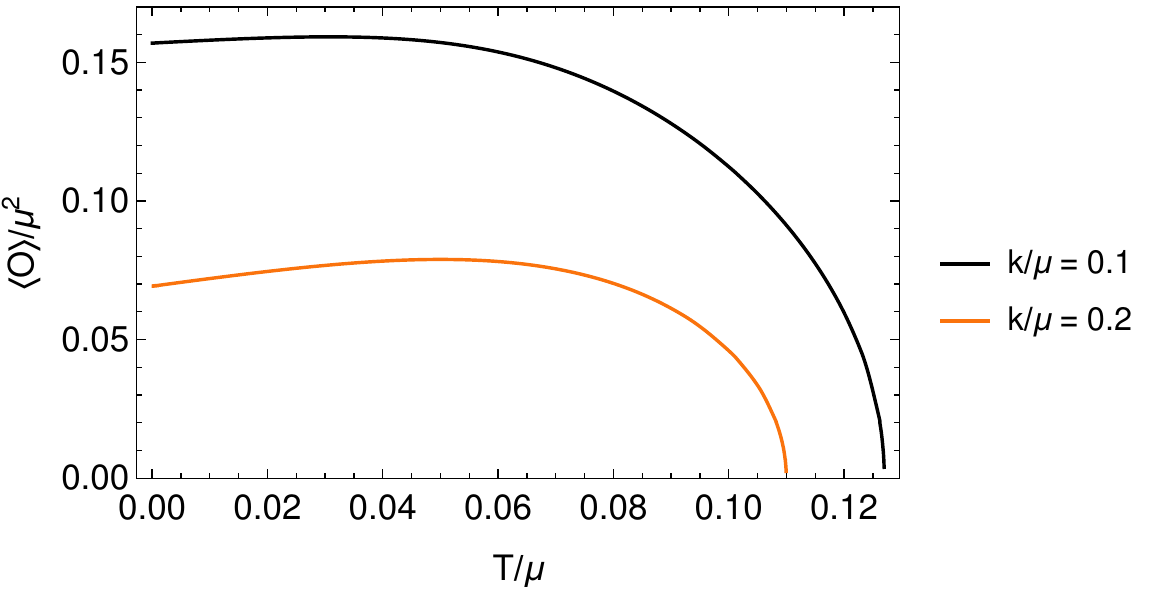}&\includegraphics[width=.45\textwidth]{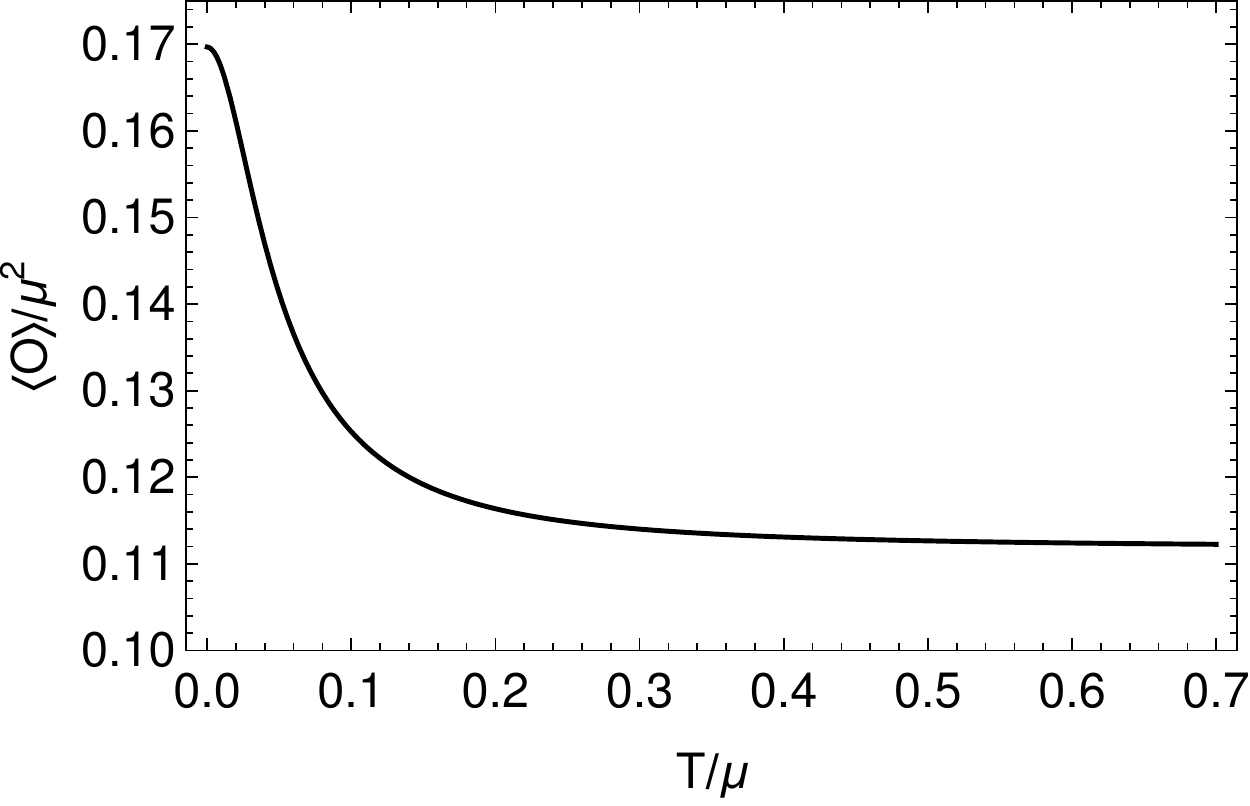}
\end{tabular}
\caption{Plots of the scalar condensate versus temperature. Left: model \eqref{DGmodel}. Right: model \eqref{model2}.}
\label{fig:condensate}
\end{figure}

The UV boundary conditions for these black holes are given in \eqref{UVexpansionbackground}. Their near-horizon expansion is as in \eqref{NearHorizon}. For each model, we construct a spontaneous solution (at $k=0.1$ for \eqref{DGmodel} and $k=0.01$ for \eqref{model2}) and display the temperature dependence of the condensate, defined as $\langle O\rangle=\phi_{(v)}$ from \eqref{eq:phiBC}, in figure \ref{fig:condensate}. Observe that the condensate for the second solution exists at all temperatures. This $k\neq0$ solution is smoothly connected to the $k=0$ solution which also has a condensate and exists at all temperatures. In this theory, the Reissner-Nordstr\"om black hole is not a solution of the classical equations of motion.

\begin{figure}
\begin{tabular}{cc}
\includegraphics[width=.45\textwidth]{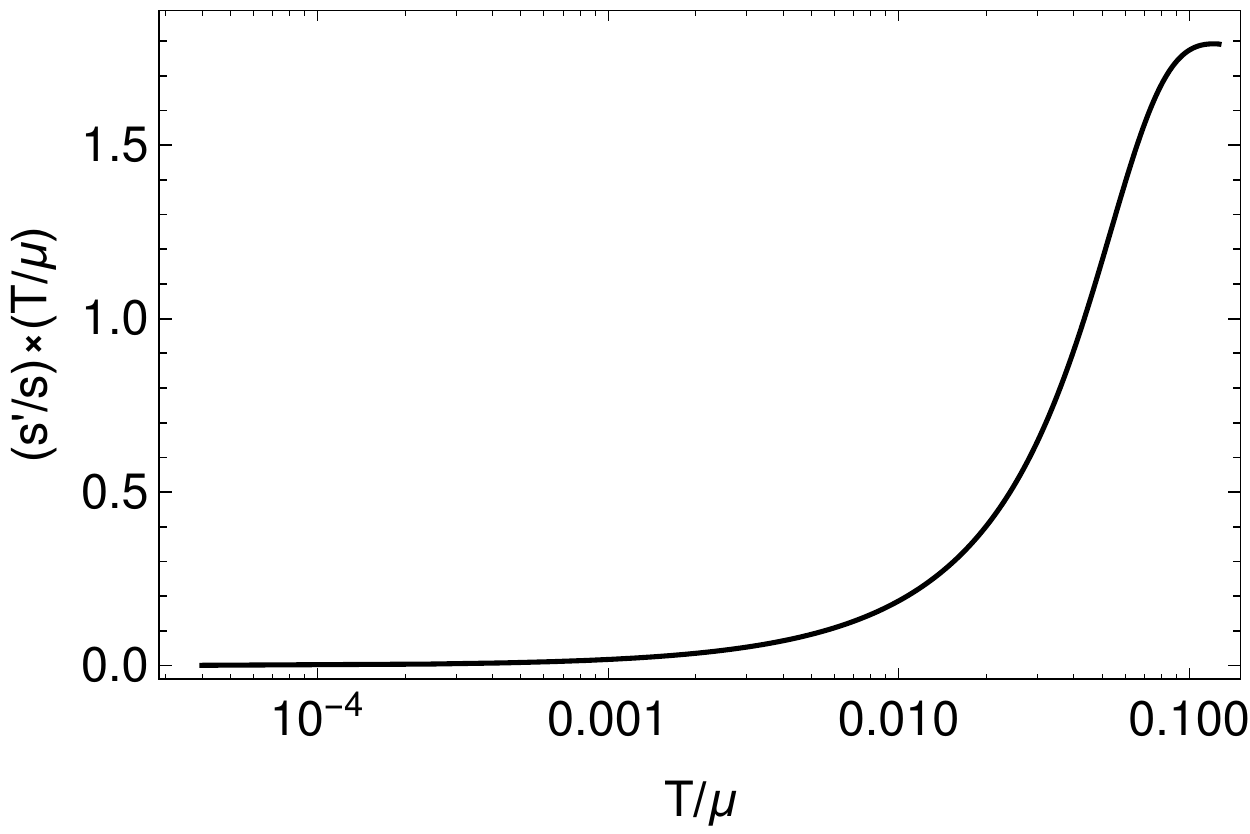}&\includegraphics[width=.45\textwidth]{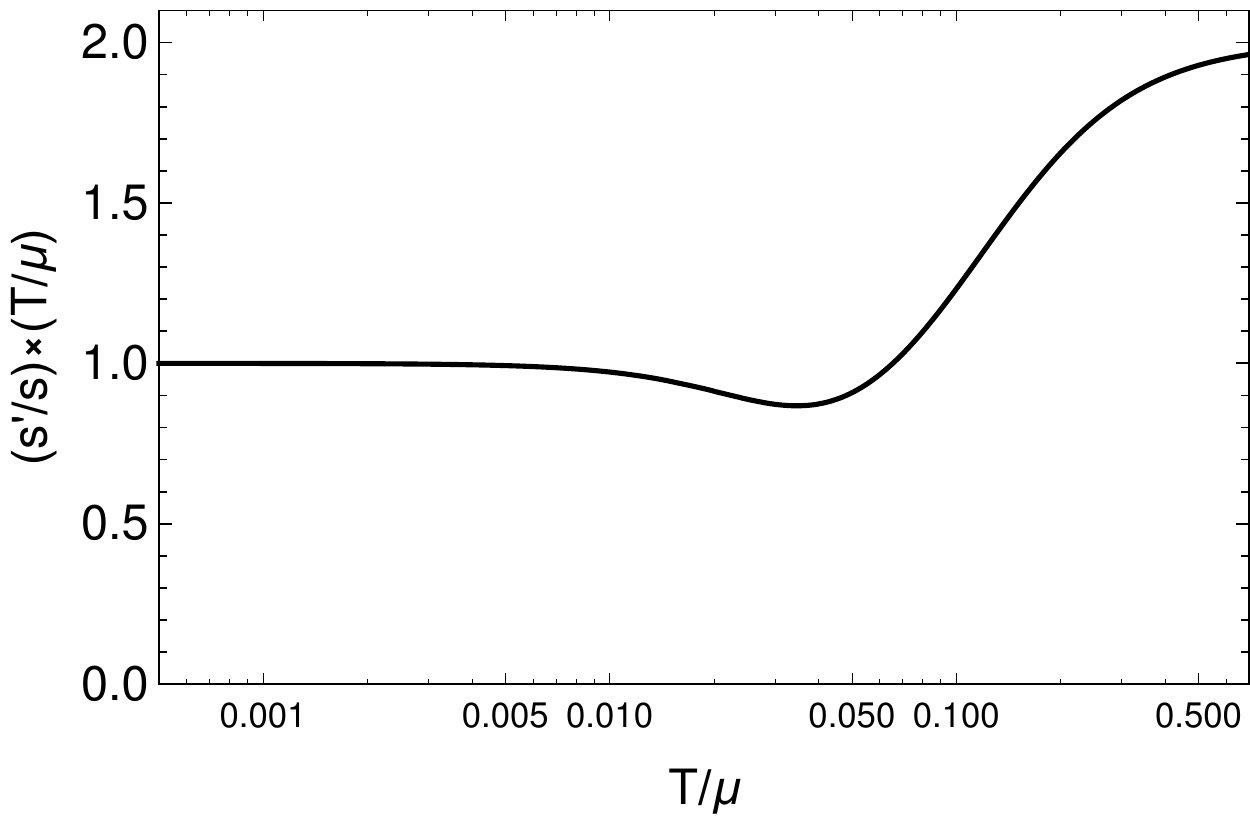}
\end{tabular}
\caption{Plots demonstrating the low temperature behavior of the entropy density. Left: the AdS$_2\times$R$^2$ zero temperature solution of the model \eqref{DGmodel} with non-vanishing entropy. Right: the semi-locally critical zero temperature solution of the model \eqref{model2} with linearly vanishing entropy.}
\label{fig:entropy}
\end{figure}

 We show the entropy density in figure \ref{fig:entropy}. The solution of the model \eqref{DGmodel} has non-vanishing zero temperature entropy and interpolates between a UV $z=1$ and an IR $z=+\infty$ (AdS$_2\times$R$^2$) fixed point. The entropy density of the solution of the model \eqref{model2} vanishes linearly with temperature and interpolates to a hyperscaling violating, semi-locally critical IR with $\tilde\theta=1$.

\begin{figure}
\begin{tabular}{cc}
\includegraphics[width=.45\textwidth]{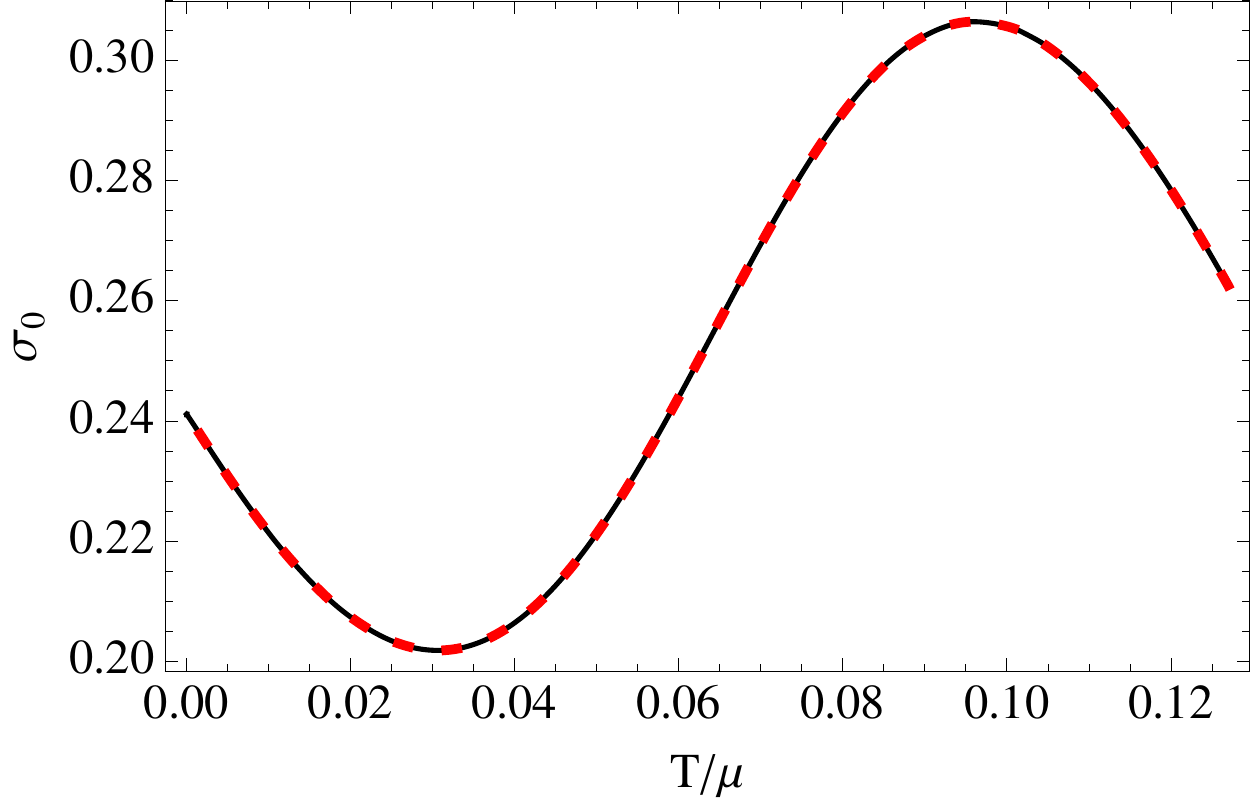}&\includegraphics[width=.45\textwidth]{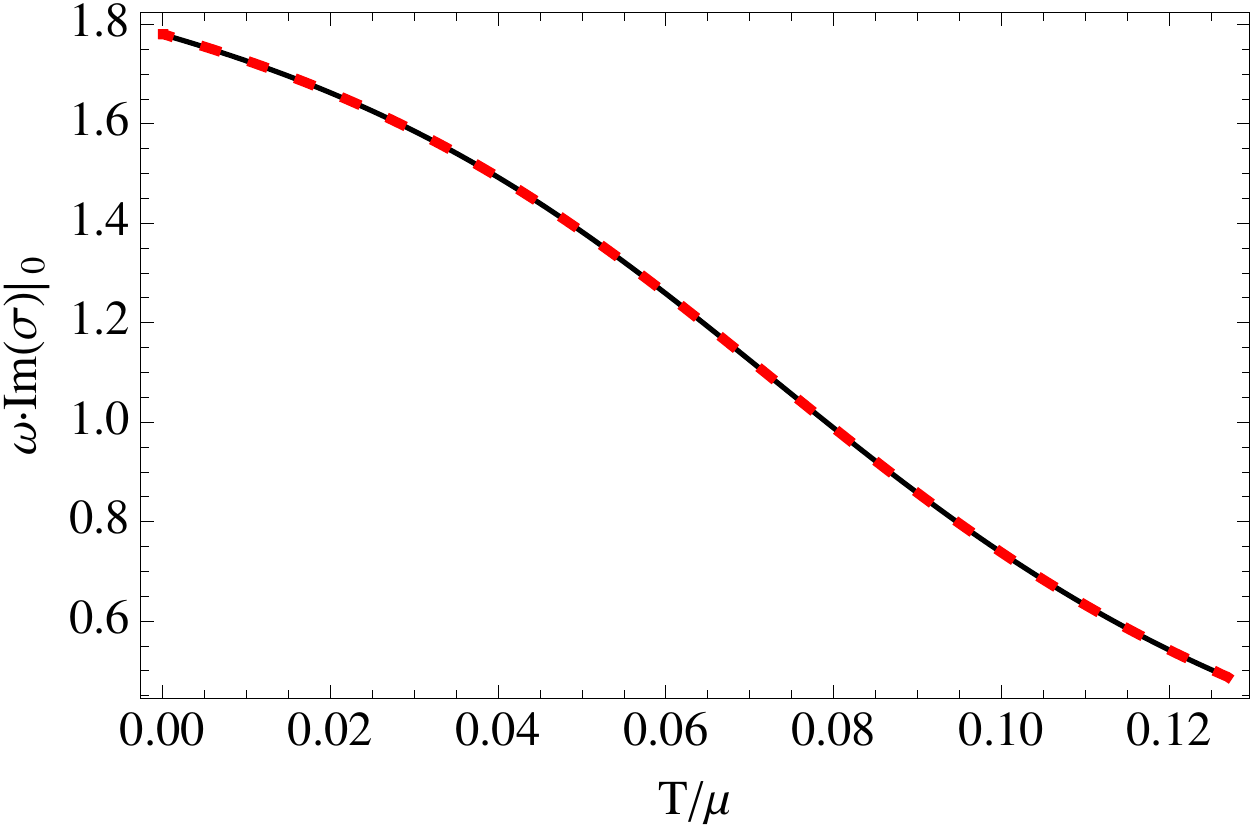}\\
\includegraphics[width=.45\textwidth]{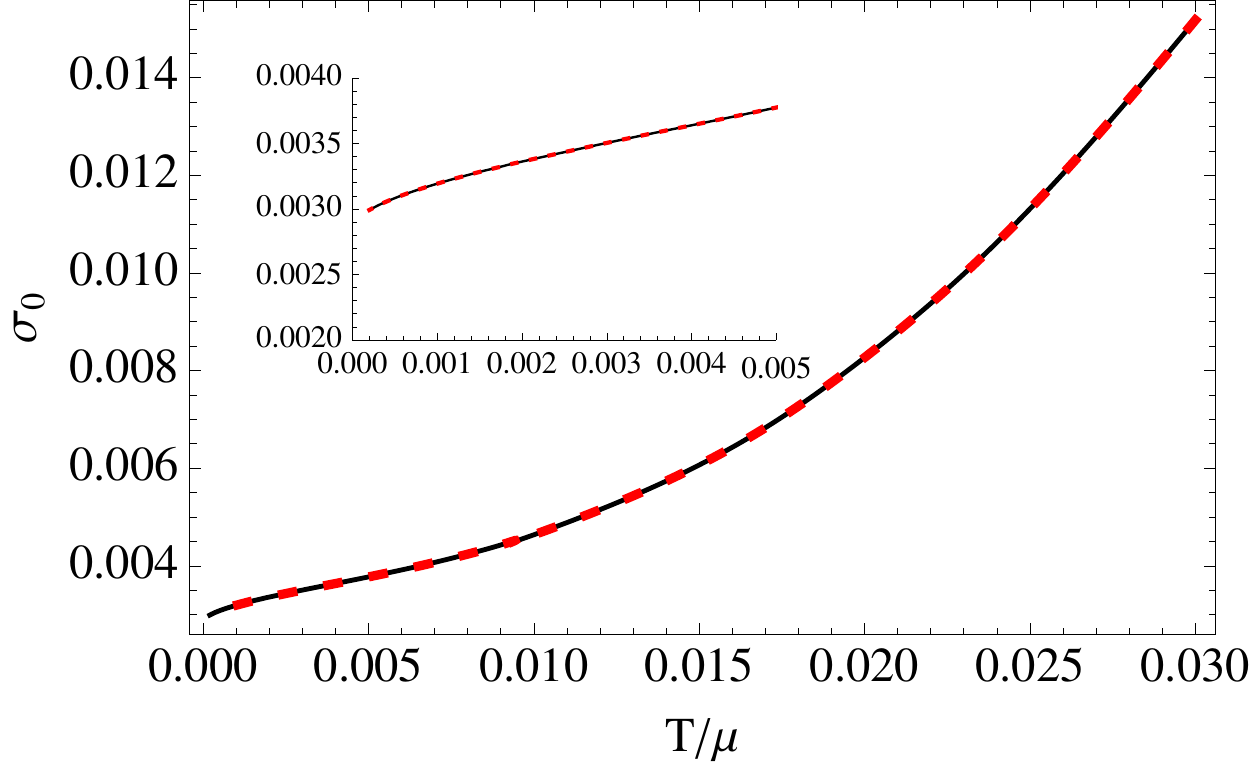}&\includegraphics[width=.45\textwidth]{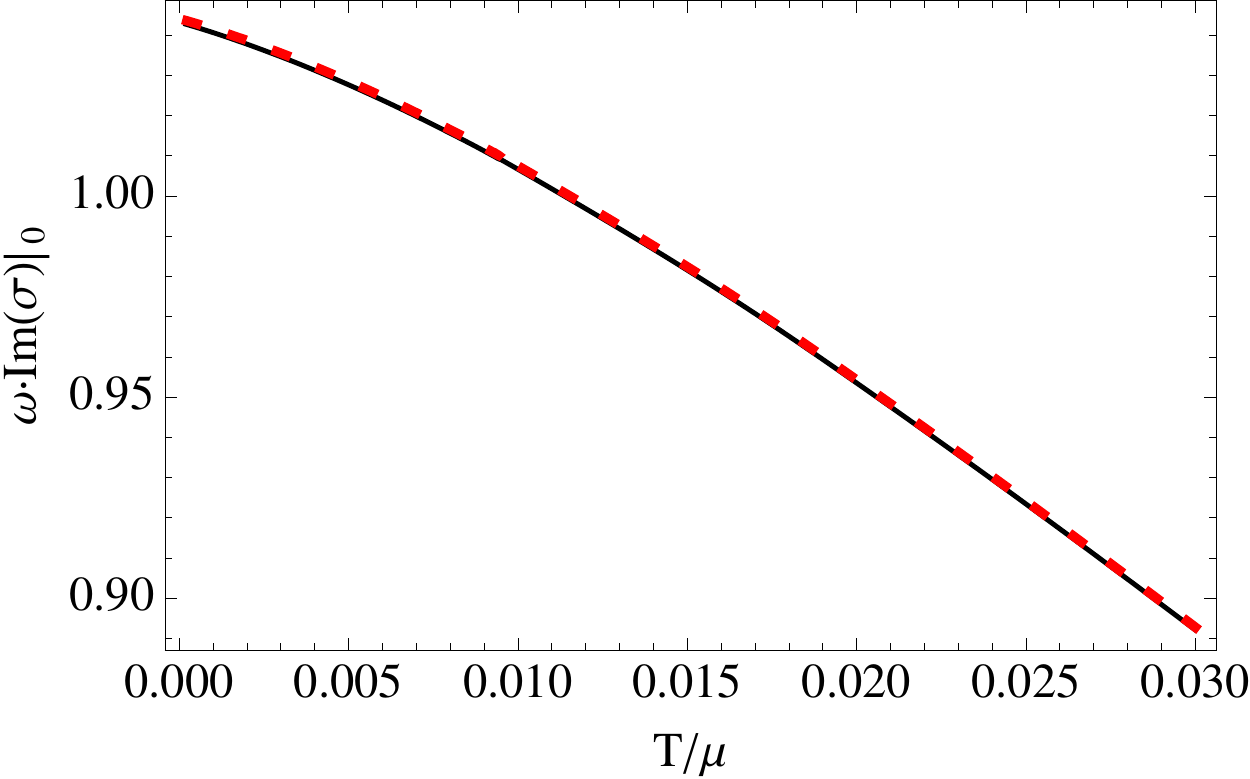}\\
\includegraphics[width=.45\textwidth]{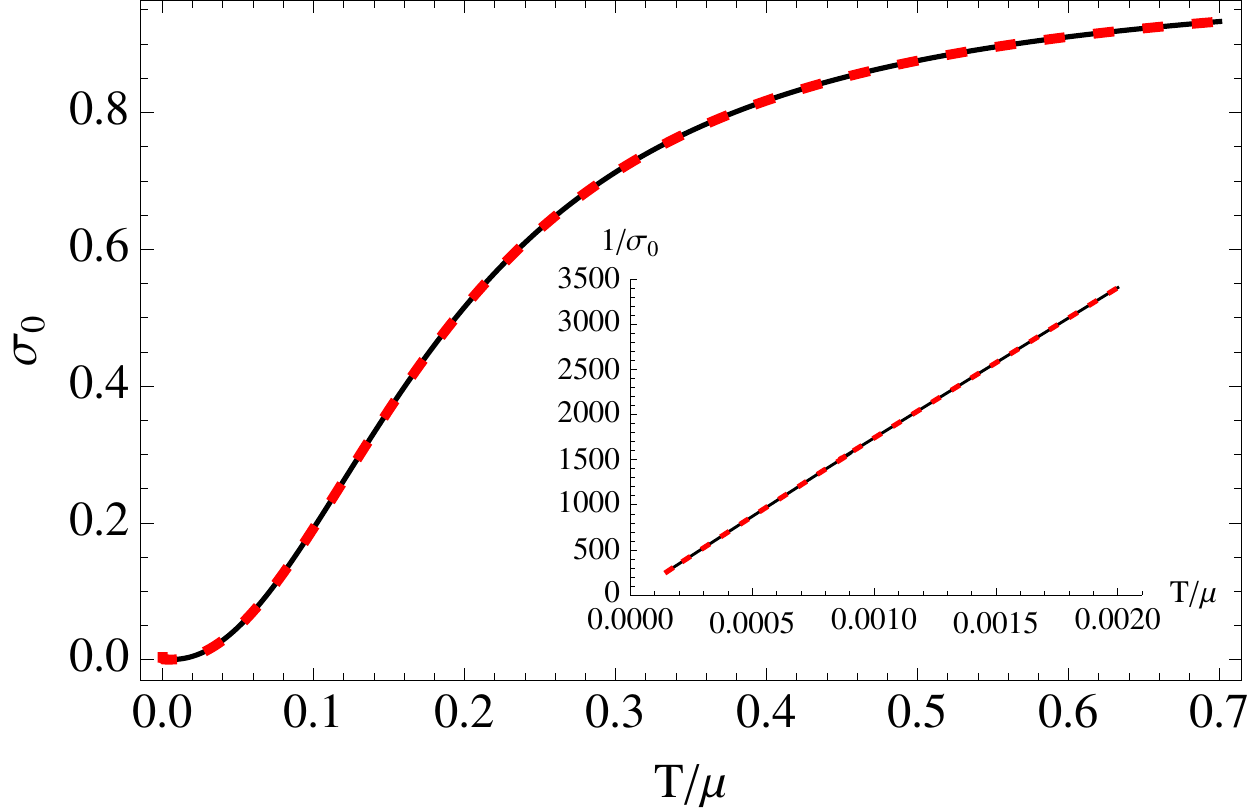}&\includegraphics[width=.45\textwidth]{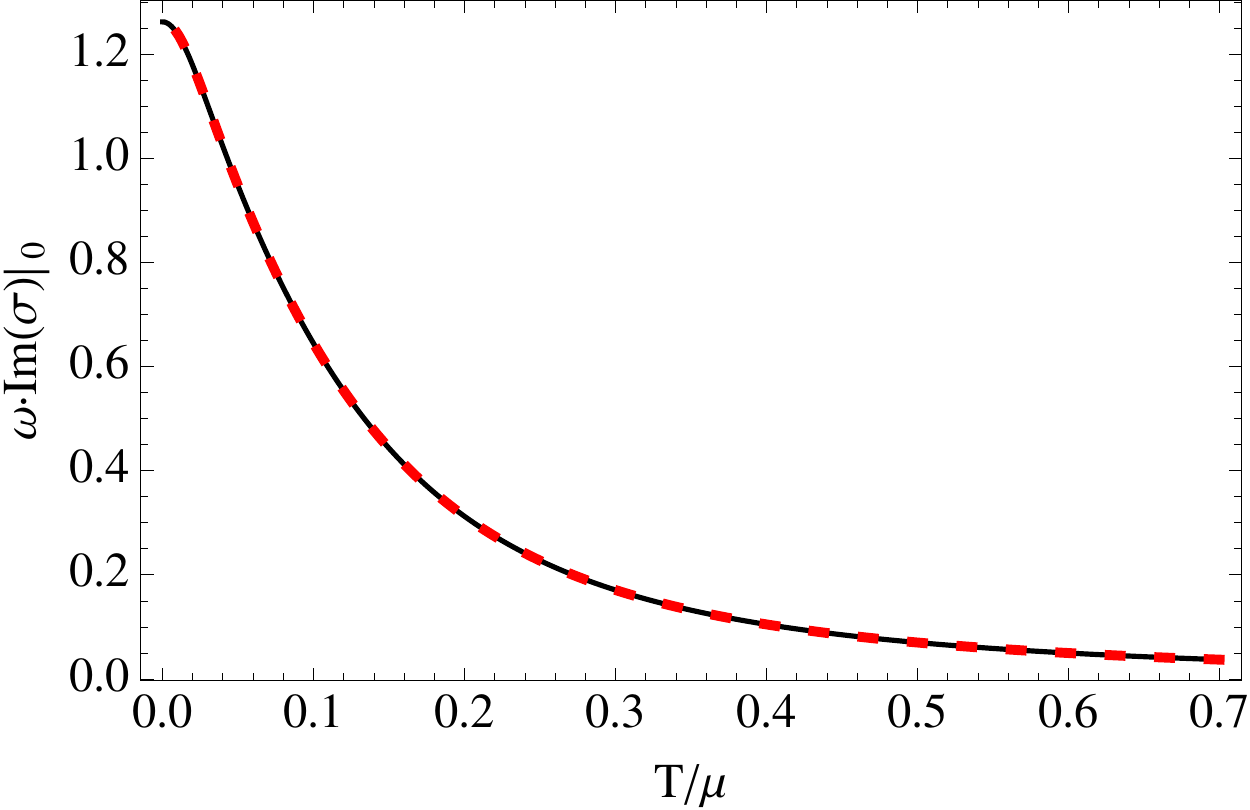}
\end{tabular}
\caption{Plots demonstrating the agreement of the numerically computed optical conductivity (red, dashed) with the analytical prediction \eqref{ACCond} (black, solid). Left: the zero frequency limit of the real part of the conductivity. Right:  The weight of the $\omega=0$ pole. Top: model \eqref{DGmodel} with $\gamma=4$; the zero temperature geometry is AdS$_2\times$R$^2$, $\sigma_o$ is non-vanishing at $T=0$ and `metallic' $\left.d\sigma_o/dT<0\right|_{T\to0}$. Middle: model \eqref{DGmodel} with $\gamma=1$; the zero temperature geometry is AdS$_2\times$R$^2$, $\sigma_o$ is non-vanishing at $T=0$ and `insulating' $\left.d\sigma_o/dT>0\right|_{T\to0}$. Bottom: model \eqref{model2}; the zero temperature geometry is conformal to AdS$_2\times$R$^2$ with $\tilde\theta=1$ and $\sigma_o$ as $1/T$ at low temperature. }
\label{fig:condeta1}
\end{figure}

Next, we compute the optical conductivity of the black holes, by perturbing the background with the Ansatz
\begin{equation}
\delta A_x(r,t)=a(r) e^{-i\omega t}\ ,\qquad h_{t}^x(r,t)=h_1(r) e^{-i\omega t}\ ,\qquad \delta \psi_x(r,t)=\chi(r) e^{-i\omega t}\ ,
\end{equation}
which is a consistent set of perturbations. In the UV, we wish to impose boundary conditions turning on an oscillating electric field. The UV expansion of the perturbation as $r\to0$ is
\begin{equation}
a=a_{(0)}+a_{(1)} r+\mathcal O(r^2)\ ,\qquad h_1=h_{(0)}+h_{(3)} r^3+\mathcal O(r^3)\ ,\qquad \chi=\frac{\chi_{(0)}}{r}+\chi_{(1)}+\mathcal O(r)\,.
\end{equation}
In contrast to \cite{Donos:2013eha}, and due to our boundary conditions \eqref{UVexpansionbackground}, we can consistently set $\chi_{(0)}=0$ (we can use a gauge transformation to set $h_{(0)}$ to zero, which shifts $\chi_{(1)}$ and not $\chi_{(0)}$). The conductivity reads
\begin{equation}
\sigma(\omega)=-\frac{i}{\omega}\frac{a_{(1)}}{a_{(0)}}\ .
\end{equation}
In figure \ref{fig:condeta1}, we show that it agrees with \eqref{ACCond}. In particular, its real part at zero frequency agrees with our analytical result \eqref{eq:sigmarealpart}, and the weight of the $\omega=0$ pole of the imaginary part is $\rho^2/\chi_{PP}$, with $\rho$ and $\chi_{PP}$ given in \eqref{density} and \eqref{chiPPdef}, respectively.


\section{Discussion and outlook \label{section:outlook}}

We have presented an effective holographic theory of CDW states. We have implemented spontaneous breaking of translations in a homogeneous manner, which corresponds to considering directly the low energy dynamics of the phonons coupled to conserved currents. At two-derivative level, our model captures excited equilibrium states with non-zero strain. By adding higher-derivative terms, we also capture thermodynamically stable phases which minimize the free energy. Our model contains quantum critical CDW zero temperature states for both cases. We computed the conductivity of these holographic CDWs, finding complete agreement between our analytic formul\ae\ and our numerics for strained phases, and with other literature \cite{Donos:2018kkm} for stable phases. As we explain in \cite{Amoretti:2017axe}, the real part of the dc conductivity may be used to predict the temperature scaling of the resistivity of CDWs with weak disorder. The zero temperature state can be insulating or metallic, depending on the details of the model. In \cite{Amoretti:2017axe}, we also connect our results with the phenomenology observed in underdoped cuprates with static charge order, and speculate on the potential relevance of the strained phases to the strange metallic region. 

We have focused only on a subset of the observables that can be computed in a state breaking translations spontaneously.  It would be worthwhile to look at the spectrum of collective excitations (eg transverse sound modes) and compare it to hydrodynamic expectations \cite{Delacretaz:2017zxd} as well as previous holographic results \cite{Argurio:2015wgr,Amoretti:2016bxs,Jokela:2017ltu,Andrade:2017cnc,Alberte:2017cch,Andrade:2017ghg}.

\cite{Andrade:2017cnc,Andrade:2017ghg} found that the low temperature resistivity of weakly-pinned holographic spatially modulated states scales with temperature. Our calculation of the incoherent conductivity, together with the quantum critical zero temperature states, could shed light on these results. More generally, it would be interesting to work out how our holographic EFT can be related to the low wavelength dynamics of inhomogeneous holographic states \cite{Donos:2017ihe}. This would provide a derivation of which higher derivative terms could arise.

It would also be interesting to revisit the analyses of commensurability effects (or lack thereof)  \cite{Andrade:2015iyf,Andrade:2017leb} in our improved model with quartic derivatives. We observed that these terms could trigger dynamical instabilities of translation-invariant phases. The higher derivative terms we consider might also inspire kinetic Mexican-hat constructions for non-holographic EFTs avoiding kinetic terms with the `wrong sign'.

Many spatially-modulated instabilities are measured in a magnetic field \cite{Wu2011,Chang2012,Ghiringhelli821}. Parity-violating spatially modulated phases have also been constructed in holography. It should be possible to extend previous holographic studies of magnetotransport of disordered metallic phases \cite{Blake:2014yla,Lucas:2015pxa,Blake:2015ina,Amoretti:2015gna,Kim:2015wba} to holographic CDW states.

\acknowledgments
We would like to thank Riccardo Argurio for collaboration at an early stage. We would like to thank Matteo Baggioli, Carlos Hoyos, Francisco Ib\'{a}\~{n}ez, Elias Kiritsis, Sasha Krikun, Nicodemo Magnoli, Alfonso Ramallo, Javier Tarr\'\i o, Paolo di Vecchia and Jan Zaanen for stimulating and insightful discussions. We are grateful to Sean Hartnoll for comments on a previous version of the manuscript. BG has been partially supported during this work by the Marie Curie International Outgoing Fellowship nr 624054 within the 7th European Community Framework Programme FP7/2007-2013. The work of D.M. was supported by grants FPA2014-52218-P from Ministerio de Econom\'\i a y Competitividad. D.A. is supported by the 
7th Framework Programme (Marie Curie Actions) under grant agreement 317089 (GATIS) from the grant CERN/FIS-NUC/0045/2015, and by the Simons Foundation grants 488637 and 488649 (Simons collaboration on the Non-perturbative bootstrap).
D.A. and D.M. thank the FRont Of pro-Galician Scientists for unconditional support. B.G. would like to thank the institute AstroParticle and Cosmology, Paris for warm hospitality at various stages of this work.

\bibliography{STSB}

\end{document}